%% file: 0usenix.tex
\documentclass[letterpaper,twocolumn,10pt]{article}
\PassOptionsToPackage{table}{xcolor}
\usepackage{usenix-2020-09}

\input{00macros}

\includeversion{arxiv}
\begin{document}
%-------------------------------------------------------------------------------

%don't want date printed
\date{}

\title{\Large \bf \titlename}

%for single author (just remove % characters)
\begin{arxiv}
\author{
{\rm Vasisht Duddu$^1$, Lipeng He$^{1}$, Asim Waheed$^{1}$, N. Asokan$^{1,2}$}\\
$^1$University of Waterloo, $^2$KTH Royal Institute of Technology\\
\{vasisht.duddu, lipeng.he, asim.waheed\}@uwaterloo.ca, asokan@acm.org
} % end author
\end{arxiv}

\maketitle

\begin{abstract}
\input{01abstract}
\end{abstract}
% \begin{arxiv}
% \begingroup\renewcommand\thefootnote{\textdagger}
% \footnotetext{See ``Statement of Contributions''}
% \endgroup
% \end{arxiv}

\input{1introduction}
\input{2background}
\input{3factors}
\input{4systematization}
\input{5evaluation}
\input{6capability}

\input{8discussions}

% \newpage
% \begin{submit}
\subsection*{Ethical Consideration}

We discuss ethical considerations based on impact to various stakeholders, risks to ethics during the research and publication process, various mitigations to reduce ethical risks, justification for conducting research and publishing.

\noindent\textbf{\underline{Impacts to Stakeholders:}}
The primary stakeholders impacted by this work are: \emph{data subjects} whose susceptibility to risks (e.g., to privacy or fairness) may worsen with colluding adversaries; \emph{model owners and service providers} who face increased security and privacy failures when threats are evaluated as interactions among risks instead of isolated risks; \emph{ML security researchers and practitioners} who may adopt our framework for auditing and designing defenses; and \emph{society} as colluding adversaries can undermine trust in ML systems used in high-stakes domains.
% \begin{itemize}[leftmargin=*]
%     \item \textbf{data subjects} whose susceptibility to risks (e.g., to privacy or fairness) may worsen with colluding adversaries.
%     \item \textbf{Model Owners and Service Providers} who face increased security and privacy failures when threats are evaluated as interactions among risks instead of isolated risks.
%     \item \textbf{ML Security Researchers and Practitioners} who may adopt our framework for auditing and designing defenses.
%     \item \textbf{Society} as colluding adversaries can undermine trust in ML systems used in high-stakes domains.
% \end{itemize}

\noindent\textbf{\underline{Risks and Benefits of Research:}}
% \noindent\textbf{Research Process.}
We do not collect new data, involve human subjects, or target real-world systems. All empirical results validate conjectures using standard datasets and models from prior work. As such, the research process itself poses minimal direct risk to data subjects or institutions.

% \noindent\textbf{Publication Impacts.} 
By explicitly characterizing train–test collusion (where training-time manipulation increases susceptibility to inference-time attacks) and test-time collusion (where outcomes of one inference attack satisfy the prerequisites of another), our framework may lower the barrier for adversaries to identify effective attack sequences.

However, the primary ethical benefit is defensive: our analysis reveals that evaluating risks in isolation can substantially underestimate real-world harm. In particular, our systematization shows that mitigations targeting a single risk may fail when adversaries collude across stages of the ML pipeline. This can help model designers account for such risks and design more effective defenses.

\noindent\textbf{\underline{Mitigations:}}
We do the following to limit potential harm:
\begin{itemize}[leftmargin=*]
\item \textbf{Using Existing Attacks:} Our framework relies on previously published risks and attacks and show how existing attacks can be amplified under collusion.
\item \textbf{Defensive Framing:} Collusions are presented as evaluation blind spots, intended to inform auditing and defense rather than to maximize adversarial effectiveness.
\item \textbf{Scoped Empirical Validation:} Experiments are limited to validating conjectured relationships between risks, not to demonstrating real-world exploitability.
\end{itemize}
There is still risk that adversaries could use our framework for attack design, but this is an inherent trade-off in our work.

\noindent\textbf{\underline{Decision to Conduct and Publish:}}
We justify publishing this work under the principles of \emph{beneficence and public interest}. Colluding adversaries already exist in practice; failing to make these interactions explicit leaves defenders unprepared and disproportionately harms data subjects. Therefore, we conclude that the benefits of publication outweigh the risks.

\subsection*{Open Science}
The code for our experiments will be made publicly available upon publication. To align with USENIX Security's open science policy, the code is available at: \url{https://github.com/ssg-research/sok-collusion}. This includes all relevant datasets, scripts, binaries, and source code. 
% After acceptance, we will share the code for artifact evaluation according to availability, functionality, and reproducibility.
% \end{submit}

\begin{arxiv}
\input{00ack}
\end{arxiv}

\bibliographystyle{plain}
\bibliography{paperS}

\input{9appendix}

\end{document}

%% file: 00macros.tex
\usepackage{versions}
\usepackage{xspace}
\usepackage[table]{xcolor}
\usepackage[inline]{enumitem}
\setlist[itemize]{itemsep=0pt, topsep=0pt, partopsep=0pt, parsep=0pt}
\usepackage{multirow}
\usepackage{tabularx}
\usepackage[flushmargin,hang]{footmisc}
\usepackage{graphics}
\usepackage{graphicx}
\usepackage{amsfonts}
\usepackage{amsmath}
\usepackage{fontawesome}
\usepackage{pifont}
\usepackage{array}
\usepackage{soul}
\usepackage[T1]{fontenc}
\usepackage{booktabs}
\usepackage{tikz}
\usepackage{amsthm}
\usepackage{mdframed}
\usepackage{thmtools}
\usepackage[most]{tcolorbox}
\usetikzlibrary{shapes,arrows}
\usetikzlibrary{tikzmark}
\usetikzlibrary{backgrounds, positioning, fit}
\PassOptionsToPackage{hyphens}{url}

\usepackage[compact]{titlesec}

\usepackage{array}
\newcolumntype{C}[1]{>{\centering\arraybackslash}p{#1}}
\newcolumntype{P}[1]{>{\centering\arraybackslash}m{#1}}

\usepackage{wasysym}
\newcommand{\minimum}{\ding{72}\xspace}%
\newcommand{\trend}{$\uparrow$\xspace}%
\newcommand{\xmark}{{\color{red}\ding{55}}\xspace}

\newcommand{\titlename}{SoK: Colluding Adversaries in Machine Learning Pipelines} %Interactions among Machine Learning Risks from 

\newcommand{\attackone}{\textbf{\texttt{Att}}$_1$\xspace}
\newcommand{\attacktwo}{\textbf{\texttt{Att}}$_2$\xspace}
\newcommand{\commonfactor}{\texttt{<f$_{c}$>}\xspace}

\newcommand{\dtrain}{\mathcal{D}_{tr}\xspace}
\newcommand{\daux}{\mathcal{D}^{aux}\xspace}
\newcommand{\dauxtrain}{\mathcal{D}^{aux}_{tr}\xspace}
\newcommand{\dauxtest}{\mathcal{D}^{aux}_{te}\xspace}
\newcommand{\dtest}{\mathcal{D}_{te}\xspace}

\newcommand{\model}{\mathcal{M}\xspace}
\newcommand{\shadowmodel}{\mathcal{M}_{shdw}\xspace}
\newcommand{\surrogatemodel}{\mathcal{M}_{sur}\xspace}

\makeatletter
\newcommand\incircbin
{%
  \mathpalette\@incircbin
}
\newcommand\@incircbin[2]
{%
  \mathbin%
  {%
    \ooalign{\hidewidth$#1#2$\hidewidth\crcr$#1\ovoid$}%
  }%
}

\makeatother

\newcommand{\threatModel}{\protect{$\textbf{\texttt{TM}}$}\xspace}

\newcommand{\adv}{\protect{$\mathcal{A}dv$}\xspace}
\newcommand{\advSet}{\protect{$\overline{\textbf{\texttt{Role}}}$}\xspace}
\newcommand{\client}{\protect{$\texttt{Clnt}$}\xspace}
\newcommand{\serviceProvider}{\protect{$\texttt{SrvProv}$}\xspace}
\newcommand{\dataOwner}{\protect{$\texttt{DtOwnr}$}\xspace}
\newcommand{\modelOwner}{\protect{$\texttt{ModOwnr}$}\xspace}
\newcommand{\modelTrainer}{\protect{$\texttt{ModTrnr}$}\xspace}
\newcommand{\modelProvider}{\protect{$\texttt{ModProv}$}\xspace}
\newcommand{\codeProvider}{\protect{$\texttt{CodeProv}$}\xspace}
\newcommand{\dataProvider}{\protect{$\texttt{DtProv}$}\xspace}

\newcommand{\knowledgeModelSet}{\protect{$\overline{\textbf{\texttt{K}}_{\textbf{\texttt{M}}}}$}\xspace}
\newcommand{\blackbox}{\protect{$\texttt{BBox}$}\xspace}
\newcommand{\graybox}{\protect{$\texttt{GBox}$}\xspace}
\newcommand{\whitebox}{\protect{$\texttt{WBox}$}\xspace}

\newcommand{\knowledgeDataSet}{\protect{$\overline{\textbf{\texttt{K}}_{\textbf{\texttt{D}}}}$}\xspace}
\newcommand{\noOverlap}{\protect{$\texttt{NoOvlap}$}\xspace}
\newcommand{\partialOverlap}{\protect{$\texttt{PartOvlap}$}\xspace}
\newcommand{\fullOverlap}{\protect{$\texttt{FullOvlap}$}\xspace}
\newcommand{\noAccess}{\protect{$\texttt{NoAccess}$}\xspace}

\newcommand{\blind}{\protect{$\texttt{Blind}$}\xspace}
\newcommand{\noBlind}{\protect{$\texttt{NoBlind}$}\xspace}

\newcommand{\capabilitySet}{\protect{$\overline{\textbf{\texttt{Cap}}}$}\xspace}
\newcommand{\passiveAdv}{\protect{$\texttt{HbC}$}\xspace}
\newcommand{\activeAdv}{\protect{$\texttt{Mal}$}\xspace}

\newcommand{\optimizationSet}{\protect{$\overline{\textbf{\texttt{Opt}}}$}\xspace}
\newcommand{\adaptive}{$\texttt{Adpt}$\xspace}
\newcommand{\nonAdaptive}{\protect{$\texttt{nonAdpt}$}\xspace}

\newcommand{\interactionSet}{$\overline{\textbf{\texttt{I}}}$\xspace}
\newcommand{\kShot}{$\texttt{KShot}$\xspace}
\newcommand{\oneShot}{$\texttt{1Shot}$\xspace}
\newcommand{\unlimited}{\protect{$\texttt{Unlmtd}$}\xspace}

\newcommand{\hardLabel}{\protect{$\texttt{hardLbl}$}\xspace}
\newcommand{\topK}{\protect{$\texttt{topK}$}\xspace}
\newcommand{\fullPrediction}{\protect{$\texttt{fullPred}$}\xspace}

\newcommand{\objectiveSet}{$\overline{\textbf{\texttt{Obj}}}$\xspace}
\newcommand{\evasion}{$\texttt{Evsn}$\xspace}
\newcommand{\poison}{$\texttt{Poisn}$\xspace}
\newcommand{\backdoor}{$\texttt{Bkdr}$\xspace}
\newcommand{\mia}{$\texttt{MemInf}$\xspace}
\newcommand{\aia}{$\texttt{AttInf}$\xspace}
\newcommand{\datarecon}{$\texttt{DtRecon}$\xspace}
\newcommand{\dia}{$\texttt{DistInf}$\xspace}
\newcommand{\modelext}{$\texttt{ModExt}$\xspace}
\newcommand{\disc}{$\texttt{Discr}$\xspace}

\definecolor{ForestGreen}{cmyk}{0.91,0,0.88,0.12}
\newcommand{\partialColl}{$\downarrow$\xspace}

\newcommand\change[1]{{\color{black}{#1}\xspace}}
\newcommand\cosemtic[1]{{\color{black}{#1}\xspace}}

\definecolor{blond}{rgb}{0.98, 0.94, 0.75}

\declaretheoremstyle[
    numbered=no, 
    headfont=\bfseries,
    postheadspace=0pt,
    headpunct={}
]{thmsty}

\declaretheorem[style=thmsty, name={}]{takeaway}

\tcolorboxenvironment{takeaway}{enhanced jigsaw, colback=blond, drop shadow, boxrule=0.9pt, boxsep=0.1pt, left=4pt, right=4pt, top=4pt, bottom=4pt}

\definecolor{warmgray}{rgb}{0.90, 0.88, 0.85}

\tcolorboxenvironment{note}{enhanced jigsaw, colback=warmgray, drop shadow, boxrule=0.9pt, boxsep=0.1pt, left=4pt, right=4pt, top=4pt, bottom=4pt}

\usetikzlibrary{shapes,arrows}
\usetikzlibrary{backgrounds, positioning, fit}
\makeatletter
\tikzset{
    database/.style={
        path picture={
            \draw (0, 1.5*\database@segmentheight) circle [x radius=\database@radius,y radius=\database@aspectratio*\database@radius];
            \draw (-\database@radius, 0.5*\database@segmentheight) arc [start angle=180,end angle=360,x radius=\database@radius, y radius=\database@aspectratio*\database@radius];
            \draw (-\database@radius,-0.5*\database@segmentheight) arc [start angle=180,end angle=360,x radius=\database@radius, y radius=\database@aspectratio*\database@radius];
            \draw (-\database@radius,1.5*\database@segmentheight) -- ++(0,-3*\database@segmentheight) arc [start angle=180,end angle=360,x radius=\database@radius, y radius=\database@aspectratio*\database@radius] -- ++(0,3*\database@segmentheight);
        },
        minimum width=2*\database@radius + \pgflinewidth,
        minimum height=3*\database@segmentheight + 2*\database@aspectratio*\database@radius + \pgflinewidth,
    },
    database segment height/.store in=\database@segmentheight,
    database radius/.store in=\database@radius,
    database aspect ratio/.store in=\database@aspectratio,
    database segment height=0.1cm,
    database radius=0.25cm,
    database aspect ratio=0.35,
}
\makeatother

\newcommand{\Snospace}{\S{}}

%% file: 01abstract.tex
Machine learning (ML) models are susceptible to various security, privacy, and fairness risks.
Adversaries with different characteristics (i.e., objectives, knowledge, and capabilities) \emph{can collude} by \change{executing one attack to amplify others.}
% exploiting one risk to amplify other risks.
Existing work lacks a systematic framework to explore collusion among adversaries, and to study the implications of the adversaries' characteristics.
We present a framework covering collusion (a) between train- and inference-time adversaries, and (b) among inference-time adversaries\footnote{\textbf{Terminology}: We hereafter use ``test-time'' instead of ``inference-time''.}.
Our framework accounts for \emph{factors enabling collusion} between adversaries.
We propose a guideline to conjecture about the potential for collusion using enabling factors.
We use it to explain prior work, conjecture about unexplored collusions, and empirically validate \change{five} such cases.
Finally, we discuss how adversaries' characteristics influence the potential for collusion.

%% file: 1introduction.tex
\section{Introduction}\label{sec:introduction}

Machine learning (ML) models are susceptible to a wide range of risks to security~\cite{SoKMLPrivSec,suriSoK,FenauxSoK,poisonSurvey2,modelextSurvey}, privacy~\cite{meminfSurvey,salem2023sok}, and fairness~\cite{mehrabi2021survey}.
\change{Prior work explored attacks to exploit vulnerabilities associated with each of these risks.}
However, adversaries with different \emph{characteristics} (i.e., objectives, knowledge, and capabilities) in the ML pipeline, either with different roles (e.g., model trainer and client) or with the same role (e.g., clients with different objectives), \emph{can collude}. 
% exploiting one risk to increase susceptibility to other risks.
% This allows the adversaries 
% to increase the effectiveness of the attacks (e.g.,~\cite{tramer2022truth}), and share costs by pooling resources. 
\change{Collusion is the explicit coordination among adversaries to execute one attack and use the outcomes to improve another (e.g.,~\cite{tramer2022truth}).}
% Such composition can help adversaries share costs by pooling resources.
\change{Collusion among adversaries with different objectives is well documented in other domains: cryptocurrency pump-and-dump schemes where one adversary inflates a stock, which can be exploited by others for profit~\cite{fu2025textsc}; and cyber-crime markets where an adversary sells compromised data for downstream attacks like identity theft~\cite{halcyon-iab}. Similar collusion within ML pipelines is feasible, and merits investigation.}

\change{For instance, consider two different adversaries: \adv{1} seeking to find an adversarial example against a target model $\model$ and \adv{2} seeking to build a shadow model of $\model$ either for some privacy attack ~\cite{liu2025amplifying}, or to steal $\model$'s functionality~\cite{orekondy2019knockoff}. \adv{1} sends multiple queries to $\model$ until it successfully finds an adversarial example~\cite{andriushchenko2020square}, usually discarding all other query-response pairs. But, they are valuable for \adv{2}. Therefore, \adv{1} can either sell the query-response data to \adv{2} or share the cost of querying $\model$. 
Hence, adversaries are incentivized to collude (a) via a marketplace for selling attack information, or (b) by sharing costs to avoid duplication of efforts.
}
%Additionally, a second adversary may guide the evasion adversary toward generating queries that are beneficial for its own attack while still satisfying the evasion adversary’s goal of finding successful adversarial examples.

Foreseeing all possible cases of collusion is challenging. A unified framework can help \emph{practitioners} and \emph{researchers} to identify new threats, design strong attacks to audit models, and design defenses against them.
Ideally, the framework should be
\begin{enumerate*}[label={(\roman*)}]
    \item \emph{comprehensive} (captures an adversaries' characteristics \emph{across multiple risks}) and
    \item \emph{extensible} (allows inclusion of additional risks and characteristics).
\end{enumerate*}
% This will allow systematic exploration of collusion potential among adversaries, and study the implications of adversary's characteristics on the potential for collusion.

We present the \emph{first} such framework that includes various factors aiding the potential for collusion. 
Our framework covers two collusion types identified from prior work:
\begin{enumerate*}[label={(\roman*)}]
\item between train- and test-time adversaries ($\rightarrow$~``\emph{train-test collusion}''), and 
\item among test-time adversaries ($\rightarrow$~``\emph{test-time collusion}'').
\end{enumerate*}
\cosemtic{For both, we observe that a successful attack changes some factors as an outcome, which can influence the effectiveness of other attacks relying on those factors. However, these factors differ between both collusion types.
For \emph{train–test collusion}, poisoning changes overfitting and memorization which can impact the effectiveness of test-time attacks that rely on those factors. Here, the framework includes the factors related to overfitting and memorization, that can be manipulated by an adversary.
For \emph{test-time collusion}, knowledge inferred from the first attack is an outcome that satisfies the prerequisites of other attacks. Here, the framework includes factors related to the adversary's knowledge.}

% Concretely, for \emph{train–test collusion}, we observe that exploiting train-time risks (e.g., data poisoning) influences overfitting and memorization, which in turn affects susceptibility to test-time risks (e.g., inference attacks).
% Thus, our framework includes factors that influence overfitting and memorization, which can be manipulated by a train-time adversary and exploited during test-time.
% For \emph{test-time collusion}, each risk requires certain \emph{prerequisites} for successful exploitation, and a successful attack yields some \emph{outcomes}, either the attack objective itself or some side information.
% These outcomes may then satisfy the prerequisites of other risks, enabling collusion.
% In this case, our framework models each risk in terms of its \emph{prerequisites} and \emph{outcomes} as the relevant factors.

Using this framework, we propose a guideline for conjecturing about collusion potential by identifying factors that are shared across attacks.
We empirically validate our conjectures for \change{five} unexplored cases of collusion.
Finally, we explore how an adversary's characteristics impact collusion potential and discuss implications for our guideline.
We claim the following contributions: we
\begin{enumerate}[leftmargin=*,itemsep=0pt,topsep=0pt]
    \item present the \emph{first framework} to systematically explore collusion potential using various \emph{factors}, and a \emph{guideline} to conjecture about potential collusion; (\Snospace\ref{sec:framework})
    \item use our guideline to \emph{explain collusion} in prior work identified from a \emph{literature survey};  (\Snospace\ref{sec:systematization})
    \item identify previously \emph{unexplored cases of collusion}, use our guideline to conjecture about them, and empirically validate \change{five} such conjectured cases; (\Snospace\ref{sec:evaluation}) and
    \item discuss the \emph{influence} of the adversary's characteristics on the collusion potential. (\Snospace\ref{sec:characteristics})
\end{enumerate}

%% file: 2background.tex
\section{Background and Related Work}\label{sec:background}

\noindent\textbf{\underline{Notations:}} An ML model ($\model_{\theta}$) is a function parameterized by $\theta$ that takes an input $x$ (e.g., images, tabular data, graphs, or text prompts) and produces $\model_{\theta}(x)$. For classification tasks, $\model_{\theta}(x)$ is a classification label, while for generative tasks, $\model_{\theta}(x)$ is an image or text. 
Hereafter, we denote $\model_{\theta}$ by simply writing $\model$.
We assume training data ($\dtrain$) follows certain distributional properties (e.g., a specific female-to-male ratio). 
We train $\model$ by updating $\theta$ to minimize the loss between $\model(x)$ and the expected output $y$.
After training, we evaluate the utility on a test dataset ($\dtest$), where we use accuracy for classifiers, and the quality of generated images (e.g., using Fréchet Inception Distance) and text (e.g., using perplexity) for generative models.
We refer to predictions, intermediate activations, $\theta$, and gradients as \emph{model observables}.

\noindent\textbf{\underline{\change{Attacks against} ML Models:}}
Following the taxonomy in Duddu et al.~\cite{duddu2024sok}, we consider an adversary (\adv) \change{executes an attack by exploiting vulnerabilities corresponding to security, privacy, and fairness risks:}
\begin{itemize}[leftmargin=*,wide,labelindent=0pt]
    \item \textbf{Evasion (\evasion)} modifies the inputs to force $\model$ to misclassify (classifiers)~\cite{SoKMLPrivSec,suriSoK}, or generate unacceptable images/text (generative models)~\cite{pham2024circumventing,yi2024jailbreak}.
    In classifiers, these modified inputs are called adversarial examples ($x_{adv} = x + \delta$) obtained by adding perturbations $\delta$ to $x$ where $\delta$ is computed to maximize $\model$'s loss. 
    Similarly, in generative models, we generate adversarial prompts but to evade safeguards against unacceptable content (or jailbreaking).

    \item\textbf{Poisoning (\poison)} trains $\model$ using malicious data records (\emph{poisons}) injected in $\dtrain$ to degrade $\model$'s utility on $\dtest$~\cite{poisonSurvey2}.
    A variant of poisoning is \textbf{backdoors (\backdoor)} where $\model$ is trained with poisons embedded with a specific pattern (aka \emph{trigger}), which maps an input to an \adv-chosen output. Backdoored inputs will generate \adv-chosen output, but clean inputs will result in normal behavior~\cite{abad2023sok,li2022backdoor,zhao2024weak}. \change{We refer to both of them as \poison for brevity.}

    \item\textbf{Unauthorized Model Ownership (\modelext)} occurs when \adv obtains either an identical copy of $\model$ or derives a surrogate model $\surrogatemodel$ from $\model$ using model extraction attacks~\cite{orekondy2019knockoff,thievesSesame,stealml,carlini2024stealing}. We focus on the latter setting.

    \item\textbf{Membership Inference (\mia)} identifies if an input was in $\dtrain$, by exploiting the difference in $\model$'s behavior on data records inside and outside $\dtrain$~\cite{meminfSurvey,duan2024membership,duanDiffMIA}. 

    \item\textbf{Attribute Inference (\aia)} identifies the value of an input's sensitive attribute, by exploiting the difference in $\model$'s behavior on inputs with different attribute values~\cite{jayaraman2022attribute}.
    In generative models, this translates to extracting personally identifiable information~\cite{lukas2023analyzing}.

    \item\textbf{Data reconstruction (\datarecon)} recovers inputs or data records from $\dtrain$ using access to $\model$~\cite{modelinvccs,genDataReconstruction,dlg,gradinvert2}. 
    For generative models, $\model$ can generate outputs which either replicate or closely resemble confidential $\dtrain$~\cite{aerni2024measuring,extractlm,extractionDiffusion}.

    \item\textbf{Distribution Inference (\dia)} identifies the distributional properties of $\dtrain$ by exploiting the difference in $\model$'s behavior when trained with different distribution values~\cite{melis2019exploiting,suri2022formalizing,suri2023dissecting,zhou2021property,propertyExistenceGAN}.
    For instance, it differentiates between $\model$ trained on $\dtrain$ with a male-to-female ratio of $0.1$ vs. $\model$ trained on $\dtrain$ with a ratio of $0.9$.

    \item\textbf{Discriminatory Behavior (\disc)}\footnote{Unlike other risks that involve an explicit \adv, \disc does not require one. 
    However, by assuming a test-time \adv who measures \disc, we can study collusion where a train-time \adv increases \disc, or a test-time \adv uses knowledge from \disc for other risks (\Snospace\ref{sec:systematization}).} is observed when $\model$ behaves differently across demographic subgroups identified by attributes such as race or gender on $\dtest$~\cite{mehrabi2021survey,gallegos2024bias,luccioni2024stable}. 
    This is measured using the difference in accuracy or false positive/negative rates across different subgroups.
\end{itemize}
\change{There are different attacks corresponding to each risk but for brevity, we use the notations for risks as attacks and} categorize them as \emph{train-time} (i.e., \poison) and \emph{test-time} (i.e., \evasion, \mia, \aia, \dia, \datarecon, \modelext, and \disc). 

\noindent\textbf{\underline{Comparison to Prior Surveys/SoKs:}}
Several surveys and SoKs study individual risks to ML models in isolation, including evasion~\cite{suriSoK,SoKMLPrivSec}, unfairness~\cite{mehrabi2021survey}, privacy~\cite{meminfSurvey}, backdoors~\cite{abad2023sok}, poisoning~\cite{poisonSurvey2}, and unauthorized model ownership~\cite{modelextSurvey}.
Salem et al.~\cite{salem2023sok} relate multiple privacy risks using game-based formulations, though without considering collusion among adversaries.
A separate line of work explores trade-offs among defenses and risks, where defenses designed for one risk may inadvertently amplify others~\cite{shokriSurvey,fioretto2022differential,yao2025sok,duddu2024sok,meng2025defender}.
% and specific defenses like differential privacy increase unfairness~\cite{fioretto2022differential,yao2025sok}. 
Other works have explored trade-offs among defenses such as fairness, interpretability, privacy, and robustness across diverse settings~\cite{fedtradeoff,ferry:sok,gittensTradeoff}, and explore how to combine defenses
% Further, prior works have explored how defenses against one risk can unintentionally affect others~\cite{duddu2024sok,meng2025defender}.
% Finally, some defenses can be combined 
without conflict to mitigate multiple risks simultaneously~\cite{duddu2024combining,szyller2023conflicting}.
% \begin{note}
Prior works either focus on systematizing individual risks or trade-offs induced by the defender's choices. This includes unintended interactions of defenses with risks (e.g., \cite{duddu2024sok,meng2025defender}) or conflicting interactions among defenses (e.g., \cite{duddu2024combining,szyller2023conflicting}).
None of the prior works systematize how adversaries may collude with each other.
% \end{note}

% \begin{note}
% Existing systematization and surveys cover  But none of them explore interactions among risks due to colluding adversaries.
% \end{note}

%% file: 3factors.tex
\section{Our Framework}\label{sec:framework}

Our goal is to develop a framework to explore \emph{collusion potential among adversaries}, and study the influence of \adv's characteristics on the collusion potential.
We identify the following requirements for an ideal framework:
\begin{enumerate*}[label={\textbf{R\arabic*}}]
    \item\label{comprehensive} \textbf{Comprehensive} (captures adversaries' characteristics \emph{across multiple risks, model types, and settings}); and
    \item\label{extensible} \textbf{Extensible} (allows inclusion of additional risks and characteristics).
\end{enumerate*}
% A framework which is comprehensive and extensible can be used for exploring the potential for collusion among adversaries.
We present our framework which includes:
\begin{enumerate*}[label={(\roman*)}]
    \item factors aiding the collusion potential among adversaries (\Snospace\ref{sec:factorsList}),
    \item relation between attack effectiveness and factors (\Snospace\ref{sec:revisitRisks}), and
    \item a guideline to conjecture about the collusion potential (\Snospace\ref{sec:guideline}).
\end{enumerate*}
We then study the influence of \adv's characteristics on collusion (\Snospace\ref{sec:characteristics}). Later in \Snospace\ref{sec:discussion}, we discuss how our framework meets \ref{comprehensive} and \ref{extensible}.

\subsection{Factors Aiding Collusion Potential}\label{sec:factorsList}

For illustration, we consider collusion among two adversaries (\adv{1} and \adv{2}), and discuss extending beyond two adversaries in \Snospace\ref{sec:discussion}. Based on prior work (\Snospace\ref{sec:systematization}), we identify two collusion types and discuss the factors aiding their potential:
\begin{itemize}[leftmargin=*]
    \item \textbf{Train-test Collusion:} \adv{1} can specifically optimize the train-time attacks, to increase the effectiveness of a test-time attack (as \attacktwo) executed by \adv{2}.
    \item \textbf{Test-time Collusion:} \adv{1} first executes a test-time attack (as \attackone), and uses the knowledge inferred to amplify another test-time attack (as \attacktwo) executed by \adv{2}.
\end{itemize}
% We below.
% and summarize them in Table~\ref{tab:factorSummary}.

% \input{tables/tab_factors}

\noindent\textbf{\underline{Train-Test Collusion:}} 
We conjecture that train-time attacks, (\poison) can be optimized to influence various factors underlying overfitting and memorization as an outcome (e.g., increase tail length of $\dtrain$'s distribution, distinguishability in model observables across subgroups or datasets). This can impact the effectiveness of test-time attacks that rely on those factors.
% We conjecture that exploiting train-time risks influences overfitting and memorization in $\model$. 
% This in turn affects susceptibility to various test-time risks that exploit the same factor.
Concretely, \adv{1} can optimize \poison to increase memorization and thereby the distinguishability between (a) training and non-training data records (to increase effectiveness of \mia~\cite{tramer2022truth}), (b) across subgroups (to increase effectiveness of \aia~\cite{tramer2022truth}), or (c) across datasets with different distributions (to increase effectiveness of \dia~\cite{chaudhari2023snap}).
Hence, our framework includes factors which influence overfitting and memorization that can be manipulated by \adv{1}, and exploited during test-time by \adv{2}.

We use factors influencing overfitting and memorization identified by Duddu et al.~\cite{duddu2024sok} who explore \change{an orthogonal problem} of how defenses impact susceptibility to unrelated risks. 
From these, we select factors that \adv{1} can control by optimizing \poison, and exclude hyperparameter-related ones\footnote{\change{For \poison, neither \dataOwner nor \dataProvider can access or modify model-related hyperparameters. Although \modelProvider may introduce architectural backdoors, model capacity should be maintained to avoid detection. Thus, \adv{1} cannot meaningfully manipulate hyperparameters without detection. For test-time attacks, \client acting as \adv{1}, does not control any hyperparameters. Therefore, we omit hyperparameter-related factors.}} (e.g., number of attributes, $\dtrain$ size, model capacity) and those which are insufficiently explored (e.g., priority of learning stable attributes, distance to decision boundary). \change{While we use the factors from Duddu et al.~\cite{duddu2024sok}, our problem requires a fundamentally different analysis and their work cannot be extended to study collusion potential:
\begin{enumerate*}[label={(\roman*)}]
\item for \poison, we focus on how \poison affects the factors instead of how the factors affect \poison as in Duddu et al.~\cite{duddu2024sok}; and
\item factors related to test-time collusion (described later) are novel.
\end{enumerate*}}
The framework finally includes the following factors:
\begin{enumerate}[label=\textbf{\texttt{Tr\arabic*}}, leftmargin=*, wide, labelindent=0pt,topsep=0pt,itemsep=0pt]
    \item[\hypertarget{tr1:tail}{\textbf{\texttt{Tr1-Tail}}}] (\textbf{Tail length of $\dtrain$'s distribution}): 
    Increasing the tail length of $\dtrain$'s distribution can increase the memorization of specific data records by $\model$, thereby increasing the effectiveness of attacks that exploit long-tailed distribution.
    \adv{1} controls this factor by injecting \poison to extend the tail of $\dtrain$'s distribution and thus, influence other attacks~\cite{duddu2024sok}.
    % \adv{1} optimizes \poison (typically outliers or rare data records) to increase the tail length of $\dtrain$'s distribution~\cite{duddu2024sok}. Such records are memorized thereby impacting $\model$'s behavior. 
    
    \item[\hypertarget{tr2:dataset}{\textbf{\texttt{Tr2-Data}}}] (\textbf{Distinguishability across datasets}): $\model$ behaves differently on data records inside and outside $\dtrain$~\cite{tramer2022truth}, or when trained on datasets with different distributions~\cite{chaudhari2023snap, chase2021property}. This distinguishability can increase the effectiveness of other attacks (e.g., \mia and \dia). \adv{1} can control this distinguishability by optimizing \poison.
    % \adv{1} optimizes \poison to increase the distinguishability in model observables across different datasets. Hence, $\model$ behaves differently on data records inside and outside $\dtrain$~\cite{tramer2022truth}, or on datasets with different distributions~\cite{chaudhari2023snap,chase2021property}.
    
    \item[\hypertarget{tr3:subgroups}{\textbf{\texttt{Tr3-Subgp}}}] (\textbf{Distinguishability across subgroups}): 
    $\model$ behaves differently on different demographic groups (e.g., race or sex)\footnote{\change{Distinguishability can be across any subgroups/classes but, we use "demographic groups” following prior work on \dia, \disc, and \aia.}}.
    This distinguishability can increase the effectiveness of other attacks (e.g., \aia, \dia, \disc).
    \adv{1} can control this distinguishability by optimizing \poison~\cite{solans2020poisoning}.
    % \adv{1} optimizes \poison to increase distinguishability in model observables across demographic subgroups~\cite{solans2020poisoning}. 
\end{enumerate} 
% For \hyperlink{tr2:dataset}{\textbf{\texttt{Tr2-Data}}} and \hyperlink{tr3:subgroups}{\textbf{\texttt{Tr3-Subgp}}}, in addition to tampering $\dtrain$ for \poison, \adv{1} can also force a similar behavior by manipulating the training process (e.g., objective function).

% By manipulating the above factors (referred to as ``train-test factors''), \adv{1} can increase the susceptibility to various test-time risks (see \Snospace\ref{sec:revisitRisks}).

\noindent\textbf{\underline{Test-time Collusion:}} 
For test-time collusion, \adv can increase their knowledge about $\model$ and $\dtrain$ as outcomes of executing an attack (e.g., distribution of $\dtrain$ from \dia, or functionally similar ``shadow model'' $\shadowmodel$\footnote{For some attacks (e.g., \evasion, \mia, \aia, \dia, \datarecon), \adv trains $\shadowmodel$ to mimic $\model$ using \adv's auxiliary data $\daux$ either independently or derived via \modelext, in which case $\shadowmodel$ is $\surrogatemodel$.} using \modelext), or some \emph{side information} (e.g., adversarial examples as \adv's auxiliary dataset $\daux$ from \evasion). This knowledge acts as a prerequisite for later attacks (e.g., knowing $\dtrain$'s distribution or $\model$'s functionality can improve the effectiveness of other attacks such as  \mia, \aia, \dia).
There is a potential for collusion if there exist a pair of attacks where the outcomes from \attackone can meet the prerequisites of \attacktwo.
The factors considered here focus on \adv’s knowledge which differ from those for train–test collusion (and those in Duddu et al.~\cite{duddu2024sok}), which do not account for \adv’s knowledge.
% The above train-test factors do not apply to test-time collusion, as both adversaries (acting as clients) may exploit the same factor, making it unclear how one risk impacts another.
% Here, we note that exploiting some risks needs satisfying \emph{prerequisites} (e.g., knowledge of $\dtrain$'s distribution, approximate functionality of $\model$). Also, a successfully exploiting some risks will result in some \emph{outcomes} (e.g., distribution of $\dtrain$ from \dia, or functionally similar $\surrogatemodel$ from \modelext), as well as \emph{side knowledge} (e.g., adversarial examples as $\daux$ from \evasion).

% We observe that test-time collusions are feasible as the outcomes from exploiting one risk meets the prerequisites for another. Our framework therefore captures each risk’s \emph{prerequisites} and \emph{outcomes} as factors:
\begin{enumerate}[label=\textbf{\texttt{Te\arabic*}}, leftmargin=*, wide, labelindent=0pt]
    \item[\hypertarget{req1:aux}{\textbf{\texttt{Te1-Aux}}}] (\textbf{Quality of \adv's $\daux$}): 
    Attacks that require training $\shadowmodel$ to mimic $\model$, or an attack model for a specific objective, benefit from having a high-quality $\daux$ that overlaps with $\dtrain$.
    % Some adversaries require $\daux$ to train $\surrogatemodel$ and attack models.
    In practice, $\daux$ typically does not overlap with $\dtrain$. If the overlap with $\dtrain$ or the quality of $\daux$ increases, and the attack transfers effectively to $\model$ (e.g.,~\cite{nasr2019comprehensive}).
    
    \item[\hypertarget{req2:dist}{\textbf{\texttt{Te2-Dist}}}]  (\textbf{Similarity of $\daux$'s distribution with $\dtrain$}): 
    % Attacks which require training $\surrogatemodel$ benefit from having $\daux$'s distribution to be similar to $\dtrain$.
    Attacks that require training $\shadowmodel$ should ensure that its functionality mimics that of $\model$ to achieve better attack transferability.
    If the distribution similarity between $\daux$ and $\dtrain$ increases, the attack transfers effectively to $\model$ (e.g.,~\cite{mlleaks}).
    
    \item[\hypertarget{req3:shadow}{\textbf{\texttt{Te3-Shdw}}}] (\textbf{Functional similarity between $\surrogatemodel$ and $\model$}): Some attacks require training $\surrogatemodel$ that mimic $\model$'s functionality. Hence, if \adv's knowledge about $\model$'s functionality increases, $\surrogatemodel$ will be more similar to $\model$, and the attack transfers effectively to $\model$ (e.g.,~\cite{mlleaks}).
    
    \item[\hypertarget{req4:architecture}{\textbf{\texttt{Te4-Arch}}}] (\textbf{Architecture similarity between $\surrogatemodel$ and $\model$}): In real-world settings, \adv may not know $\model$'s architecture and as the knowledge increases, $\surrogatemodel$ will be similar to $\model$ and the attack transfers effectively to $\model$ (e.g.,~\cite{mlleaks}).
\end{enumerate}
% Table~\ref{tab:factorSummary} summarizes factors enabling both the train-test and test-time collusion.
% Satisfying the above four factors will allow \adv to better approximate $\model$'s behavior, and design more effective attacks that transfer to $\model$. 

\subsection{Relation between Attacks and Factors}\label{sec:revisitRisks}
We will now describe the relation between each attack and the various factors for train-test and test-time collusion:
\cosemtic{\begin{enumerate*}[label={(\roman*)}]
\item how attack effectiveness leads to change in factors; and
\item how changes in factors correlate with attack effectiveness.
\end{enumerate*}}
\cosemtic{\begin{itemize}[leftmargin=*]
\item \textbf{Train-test Collusion:}
Table~\ref{tab:train_test_relation} shows (i) how the effectiveness of the train-time attack (i.e., \poison) leads to change in factors (left half), and (ii) how changes in factors correlate with the effectiveness of various attacks (right half).
On the other hand, the correlation between changes in the factors and the effectiveness of various attacks is reproduced from Duddu et al.~\cite{duddu2024sok} (Table~\ref{tab:train_test_relation}: right half).

\item \noindent\textbf{Test-time Collusion:} Table~\ref{tab:test_time_relation} shows (i) how the effectiveness of test-time attacks leads to changes in factors (top), and (ii) how changes in factors correlate with the effectiveness of other test-time attacks (bottom).
\end{itemize}}
In both collusion types, we use $\uparrow$ (or $\downarrow$) to indicate positive (negative) correlation (Tables~\ref{tab:train_test_relation} and~\ref{tab:test_time_relation}). 
% Refer to Table~\ref{tab:train_test_relation} for train-inference collusion, and Table~\ref{tab:test_time_relation} for inference-time collusion. 
For each attack, we describe the relation between the attacks and the enabling factors for train-test and test-time collusion below.

\input{tables/tab_factors_full}

\noindent\underline{\textbf{Poisoning} (\poison)}\\
% These risks are applicable for train-test collusions, and do not have test-time factors. We indicate \emph{how the effectiveness of \poison correlates with various factors}.
This is a train-time attack, and \cosemtic{hence, we only discuss the factors enabling train-test collusion. Below, we indicate how the effectiveness of \poison correlates with a change in the factors (in Table~\ref{tab:train_test_relation}: left half under ``Correlation of attack effectiveness with change in factors'').}
\begin{itemize}[leftmargin=*,wide,labelindent=0pt]
    \item\textbf{\emph{Train-Test Factors:}} \adv increases the tail length of $\dtrain$'s distribution by adding poisons and backdoors (\hyperlink{tr1:tail}{\textbf{\texttt{Tr1-Tail}}}: $\uparrow$)~\cite{yifeng2025swallowing}.
    The attack can be optimized to increase the distinguishability of model observables between (a) data records inside and outside $\dtrain$ (\hyperlink{tr2:dataset}{\textbf{\texttt{Tr2-Data}}}: $\uparrow$)~\cite{tramer2022truth}, (b) different distributional properties (\hyperlink{tr2:dataset}{\textbf{\texttt{Tr2-Data}}}: $\uparrow$)~\cite{chaudhari2023snap,chase2021property}, and (c) demographic subgroups (\hyperlink{tr3:subgroups}{\textbf{\texttt{Tr3-Subgp}}}: $\uparrow$)~\cite{solans2020poisoning}.
\end{itemize}

\cosemtic{For the remaining attacks, in the case of train–test collusion, we indicate how changes in the factors (due to \poison) correlate with attack effectiveness (Table~\ref{tab:train_test_relation}: right half).
For test-time collusion, we indicate both: \begin{enumerate*}[label={(\roman*)}]
\item the correlation between attack effectiveness and changes in the factors (Table~\ref{tab:test_time_relation}: top); and
\item the correlation between changes in the factors and attack effectiveness (Table~\ref{tab:test_time_relation}: bottom).
\end{enumerate*}}

\noindent\underline{\textbf{Evasion} (\evasion)}
\begin{itemize}[leftmargin=*,wide,labelindent=0pt]
    \item \textbf{\emph{Train-Test Factors:}} As the tail length of $\dtrain$ increases, models are more susceptible to adversarial examples particularly from the tail classes (\hyperlink{tr1:tail}{\textbf{\texttt{Tr1-Tail}}}: $\uparrow$)~\cite{yue2024revisiting,cho2025longtailed}. 
    Models perform well on well-represented (head) classes, but struggle to classify under-represented (tail) classes.
    
    \item \textbf{\emph{Test-time Factors:}} In a blackbox setting where \adv is required to train $\surrogatemodel$, the effectiveness of transferable adversarial examples is positively correlated with quality of $\daux$ (\hyperlink{req1:aux}{\textbf{\texttt{Te1-Aux}}}: $\uparrow$), similarity of $\daux$'s distribution with $\dtrain$ (\hyperlink{req2:dist}{\textbf{\texttt{Te2-Dist}}}: $\uparrow$), and similarity of $\surrogatemodel$'s functionality and architecture with $\model$ (\hyperlink{req3:shadow}{\textbf{\texttt{Te3-Shdw}}}: $\uparrow$, \hyperlink{req4:architecture}{\textbf{\texttt{Te4-Arch}}}: $\uparrow$).
    
    Successful \evasion provides adversarial examples as \emph{side knowledge} that can be used as $\daux$ for other attacks (\hyperlink{req1:aux}{\textbf{\texttt{Te1-Aux}}}: $\uparrow$)~\cite{prada,choquetteChoo21a,canaryMIA}. Increasing the effectiveness of \evasion results in a better quality $\daux$. 
\end{itemize}

\noindent\underline{\textbf{Membership Test} (\mia)}
\begin{itemize}[leftmargin=*,wide,labelindent=0pt]
    \item\textbf{\emph{Train-Test Factors:}}
    As the tail length increases, outlier data records (from the tail classes) are memorized more, resulting in higher susceptibility to \mia (\hyperlink{tr1:tail}{\textbf{\texttt{Tr1-Tail}}}: $\uparrow$)~\cite{carlini2022membership,carlini2022the}.
    Furthermore, susceptibility to \mia increases with higher distinguishability in model observables across data records inside and outside $\dtrain$ (\hyperlink{tr2:dataset}{\textbf{\texttt{Tr2-Data}}}: $\uparrow$)~\cite{shokri2017membership}.

    \item\textbf{\emph{Test-time Factors:}} The effectiveness of \mia is positively correlated with quality of $\daux$ (\hyperlink{req1:aux}{\textbf{\texttt{Te1-Aux}}}: $\uparrow$), similarity of $\daux$'s distribution with $\dtrain$ (\hyperlink{req2:dist}{\textbf{\texttt{Te2-Dist}}}: $\uparrow$), and similarity of $\surrogatemodel$'s functionality and architecture with $\model$ (\hyperlink{req3:shadow}{\textbf{\texttt{Te3-Shdw}}}, \hyperlink{req4:architecture}{\textbf{\texttt{Te4-Arch}}}: $\uparrow$)~\cite{shokri2017membership}.

    After a successful attack, \adv identifies data records in $\dtrain$, which can act as $\daux$ for other attacks (\hyperlink{req1:aux}{\textbf{\texttt{Te1-Aux}}}: $\uparrow$). \adv can accurately infer data records in $\dtrain$ as the effectiveness of \mia increases.
\end{itemize}

\noindent\underline{\textbf{Attribute Test} (\aia)}
\begin{itemize}[leftmargin=*,wide,labelindent=0pt]
    \item\textbf{\emph{Train-Test Factors:}}
        \aia is more effective as $\dtrain$'s tail length increases (\hyperlink{tr1:tail}{\textbf{\texttt{Tr1-Tail}}}: $\uparrow$)~\cite{jayaraman2022attribute}.
        \aia exploits the distinguishability in model observables across subgroups. Hence, increasing distinguishability will also make \aia more effective (\hyperlink{tr3:subgroups}{\textbf{\texttt{Tr3-Subgp}}}: $\uparrow$)~\cite{aalmoes2025alignment}.

    \item\textbf{\emph{Test-time Factors:}} The effectiveness of \aia is positively correlated with quality of $\daux$ (\hyperlink{req1:aux}{\textbf{\texttt{Te1-Aux}}}: $\uparrow$), similarity of $\daux$'s distribution with $\dtrain$ (\hyperlink{req2:dist}{\textbf{\texttt{Te2-Dist}}}: $\uparrow$), and similarity of $\surrogatemodel$'s functionality and architecture with $\model$ (\hyperlink{req3:shadow}{\textbf{\texttt{Te3-Shdw}}}, \hyperlink{req4:architecture}{\textbf{\texttt{Te4-Arch}}}: $\uparrow$).
        \aia can help estimate $\dtrain$'s distribution when evaluated across several data records sampled i.i.d from $\dtrain$ (\hyperlink{req2:dist}{\textbf{\texttt{Te2-Dist}}}: $\uparrow$).
The estimate about $\dtrain$'s distribution is better with more effective \aia.
\end{itemize}

\noindent\underline{\textbf{Distribution Test} (\dia)}
\begin{itemize}[leftmargin=*,wide,labelindent=0pt]
    \item\textbf{\emph{Train-Test Factors:}}
    As the tail length increases, outlier data records (from the tail classes) are memorized more, resulting in higher susceptibility to \dia (\hyperlink{tr1:tail}{\textbf{\texttt{Tr1-Tail}}}: $\uparrow$) ~\cite{chase2021property,chaudhari2023snap}.
    \dia exploits the distinguishability of model observables when trained on datasets with different properties. Hence, increasing the distinguishability will also make \dia effective (\hyperlink{tr2:dataset}{\textbf{\texttt{Tr2-Data}}}: $\uparrow$)~\cite{suri2022formalizing}.

    \item\textbf{\emph{Test-time Factors:}} The effectiveness of \dia is positively correlated with quality of $\daux$ (\hyperlink{req1:aux}{\textbf{\texttt{Te1-Aux}}}: $\uparrow$), and similarity of $\surrogatemodel$'s functionality and architecture with $\model$ (\hyperlink{req3:shadow}{\textbf{\texttt{Te3-Shdw}}}, \hyperlink{req4:architecture}{\textbf{\texttt{Te4-Arch}}}: $\uparrow$).

    After a successful attack, \adv learns $\dtrain$'s distribution (\hyperlink{req2:dist}{\textbf{\texttt{Te2-Dist}}}: $\uparrow$). The estimate about $\dtrain$'s distribution is better with more effective \dia.
\end{itemize}

\noindent\underline{\textbf{Data Reconstruction} (\datarecon)}
\begin{itemize}[leftmargin=*,wide,labelindent=0pt]
    \item\textbf{\emph{Train-Test Factors:}}
    Data records from tail classes are more susceptible (\hyperlink{tr1:tail}{\textbf{\texttt{Tr1-Tail}}}: $\uparrow$)~\cite{li2025head}.
    Higher distinguishability across datasets and subgroups results in better attack effectiveness (\hyperlink{tr2:dataset}{\textbf{\texttt{Tr2-Data}}}, \hyperlink{tr3:subgroups}{\textbf{\texttt{Tr3-Subgp}}}: $\uparrow$)~\cite{ye2024leaveoneout,duddu2024sok}.

    \item\textbf{\emph{Test-time Factors:}} 
    The effectiveness of \datarecon is positively correlated with quality of $\daux$ (\hyperlink{req1:aux}{\textbf{\texttt{Te1-Aux}}}: $\uparrow$), similarity of $\daux$'s distribution with $\dtrain$ (\hyperlink{req2:dist}{\textbf{\texttt{Te2-Dist}}}: $\uparrow$), and similarity of $\surrogatemodel$'s functionality and architecture with $\model$ (\hyperlink{req3:shadow}{\textbf{\texttt{Te3-Shdw}}}, \hyperlink{req4:architecture}{\textbf{\texttt{Te4-Arch}}}: $\uparrow$).

    After a successful attack, \adv infers data records in $\dtrain$, which can act as $\daux$ for other attacks (\hyperlink{req1:aux}{\textbf{\texttt{Te1-Aux}}}: $\uparrow$).
    \adv can more accurately infer data records in $\dtrain$ as the effectiveness of \datarecon increases.
\end{itemize}

\noindent\underline{\textbf{Unauthorized Model Ownership} (\modelext)}
\begin{itemize}[leftmargin=*,wide,labelindent=0pt]
    \item\textbf{\emph{Train-Test Factors:}}
    As the tail length increases, $\model$ may not sufficiently capture the tail classes. Hence, training $\surrogatemodel$ from $\model$ will exacerbate the utility drop on tail classes for $\surrogatemodel$ (\hyperlink{tr1:tail}{\textbf{\texttt{Tr1-Tail}}}: $\downarrow$)~\cite{same}.

    \item\textbf{\emph{Test-time Factors:}} 
    The effectiveness of \modelext is positively correlated with quality of $\daux$ (\hyperlink{req1:aux}{\textbf{\texttt{Te1-Aux}}}: $\uparrow$), and similarity of $\daux$'s distribution with $\dtrain$ (\hyperlink{req2:dist}{\textbf{\texttt{Te2-Dist}}}: $\uparrow$).

    \adv can either learn $\model$'s architecture (\hyperlink{req4:architecture}{\textbf{\texttt{Te4-Arch}}}: $\uparrow$)~\cite{wang2018stealing,joon2018towards}, or obtain a functionally similar $\surrogatemodel$ (\hyperlink{req3:shadow}{\textbf{\texttt{Te3-Shdw}}}: $\uparrow$)~\cite{orekondy2019knockoff,prada}.
    Effective attacks extract architecture and functionality more accurately.
\end{itemize}

\noindent\underline{\textbf{Discriminatory Behavior} (\disc)}
% \begin{itemize}[leftmargin=*,wide,labelindent=0pt]
%     \item\textbf{\emph{Train-Test Factors:}} As the tail length increases (addition of new minority subgroups), the effectiveness of \disc also increases (\hyperlink{tr1:tail}{\textbf{\texttt{Tr1-Tail}}}: $\uparrow$)~\cite{mehrabi2021survey}.
%     Since \disc measures distinguishability across subgroups, higher distinguishability implies more effective \disc (\hyperlink{tr3:subgroups}{\textbf{\texttt{Tr3-Subgp}}}: $\uparrow$)~\cite{aalmoes2025alignment}.
%     \item\textbf{\emph{Test-time Factors:}} \disc has no prerequisites, as it is evaluated on $\model$ using $\dtest$.
%     However, \disc can be useful to estimate $\dtrain$'s distribution (\hyperlink{req2:dist}{\textbf{\texttt{Te2-Dist}}}: \colorbox{orange!25}{$\uparrow$}).
% \end{itemize}
\change{In train-test collusion, \poison can be used to increase \disc. Here, we treat \disc (a risk without attacks) as \attacktwo to study such collusion. In test-time collusion, there are no attacks corresponding to \disc to obtain outcomes as \attackone. Thus, we omit it}.
% we adopt this assumption for uniformity and to study collusion scenarios in which executing train-time attack (i.e., \poison) increases \disc. Accordingly, we consider only train–test collusion and exclude test-time collusion, where there is no notion of ``\adv executing attacks to exploit \disc''.}
\begin{itemize}[leftmargin=*,wide,labelindent=0pt]
    \item\textbf{\emph{Train-Test Collusion:}} As the tail length increases (addition of new minority groups), the effectiveness of \disc also increases (\hyperlink{tr2:dataset}{\textbf{\texttt{Tr2-Data}}}: $\uparrow$)~\cite{mehrabi2021survey}.
    Since \disc measures distinguishability across demographic groups, higher distinguishability implies higher \disc (\hyperlink{tr3:subgroups}{\textbf{\texttt{Tr3-Subgp}}}: $\uparrow$)~\cite{aalmoes2025alignment}.

    % \item\textbf{\emph{Test-time Collusion:}} \disc has no prerequisites, as it is evaluated on $\model$ using \gls{dtest}. Upsampling or downsampling of data records in $\dtrain$ reduces discriminatory behavior~\cite{Pias202}. Hence, $\model$ with high \disc is likely to better reflect $\dtrain$'s distribution (\hyperlink{req2:dist}{\textbf{\texttt{Te2-Dist}}}: $\uparrow$)~\cite{suri2023dissecting}.
    % For instance, some fairness metrics (e.g., disparate impact) can capture this imbalance in $\dtrain$'s distribution.
\end{itemize}

\subsection{Guideline to Explain/Predict Collusion}\label{sec:guideline}

\cosemtic{We conjecture that the potential for collusion can be predicted by analyzing the common factors between the two attacks, i.e., comparing ``How attack effectiveness leads to change in factors'' and under ``How changes in factors correlate with attack effectiveness'' in Tables~\ref{tab:train_test_relation} and~\ref{tab:test_time_relation}.
However, defining an exact algorithm is challenging due to the complex and only partially understood dynamics of various factors.
Therefore, we present a guideline using expert insight to predict if there is potential for collusion.}

\cosemtic{We consider two adversaries, \adv{1} and \adv{2}, executing \attackone and \attacktwo, respectively. In train-test collusion, \attackone is always \poison and \attacktwo is a test-time attack; in test-time collusion, both are test-time attacks.}
\begin{enumerate}[leftmargin=*, label={\textbf{G\arabic*}},wide,labelindent=0pt,topsep=0pt,itemsep=0pt]

\item\label{guideline:G1} \cosemtic{Identify the common factor(s) \commonfactor by matching the non-empty entries across the two halves of Tables~\ref{tab:train_test_relation} and~\ref{tab:test_time_relation}: compare columns where the row corresponds to \attackone (under ``How attack effectiveness leads to change in factors'') with the same columns where the row corresponds to \attacktwo (under ``How changes in factors correlate with attack effectiveness'').
See Tables~\ref{tab:train_test_relation} and~\ref{tab:test_time_relation} for train–test and test-time collusion.}

\item \cosemtic{If \commonfactor is available, indicate the common factors using a pair of arrows in the format ``\commonfactor: (<$\cdot$>$_1$, <$\cdot$>$_2$)'' where <$\cdot$>$_1$ corresponds to the arrow depicting ``How attack effectiveness leads to change in factors'' (for \attackone) while <$\cdot$>$_2$ corresponds to the arrow depicting ``How changes in factors correlate with attack effectiveness'' (for \attacktwo).}
\begin{itemize}[leftmargin=*,topsep=0pt,itemsep=0pt]
\item If arrows are aligned ($\uparrow$, $\uparrow$), it suggests that \adv{1} and \adv{2} can potentially collude as executing \attackone increases \attacktwo.
\item If arrows are not aligned (e.g., $\uparrow$, $\downarrow$), it suggests \adv{1} and \adv{2} are unlikely to collude (e.g., exploiting \attackone either decreases \attacktwo or does not meet prerequisites of \attacktwo).
\end{itemize}
\item If \commonfactor is not available, collusion cannot be concluded.
\end{enumerate}
We use the guideline to explain collusion in prior work (\Snospace\ref{sec:systematization}), conjecture about unexplored ones, and empirically validate \cosemtic{five cases} (\Snospace\ref{sec:evaluation}).
We later discuss accounting for \adv's characteristics (\Snospace\ref{sec:characteristics}), and extending beyond two adversaries (\Snospace\ref{sec:discussion}).

%% file: tables/tab_factors_full.tex
\setlength{\tabcolsep}{2pt}
\begin{table}[ht]
\centering
% \begin{center}
\caption{\cosemtic{\textbf{(Train-test Collusion) Relation of attacks and factors:}
Rows for \poison under ``How effectiveness of attack (\poison) leads to change in factors'' (left half) indicates how \poison influences a change in factors, instead of how change in factors influence effectiveness of \poison (as in \cite{duddu2024sok}).
Rows under ``How changes in factors correlate with attack effectiveness'' (right half) are reproduced from \cite{duddu2024sok}; only the factors that can be manipulated by \adv are considered. $\uparrow$ (or $\downarrow$) indicate positive (negative) correlation; \colorbox{gray!20}{Gray} $\rightarrow$ not applicable; empty cells $\rightarrow$ no prior literature reported the correlations.}}
\label{tab:train_test_relation}
\vspace{0.25cm}
\resizebox{\columnwidth}{!}{
\begin{tabular}{l||P{1.5cm}|P{1.5cm}|P{1.5cm}||P{1.5cm}|P{1.5cm}|P{1.5cm}}
\bottomrule

\toprule
 & \multicolumn{3}{c||}{\textbf{How effectiveness of attack (\poison)}} & 
\multicolumn{3}{c}{\textbf{How changes in factors}}\\
& \multicolumn{3}{c||}{\textbf{leads to change in factors}} & 
\multicolumn{3}{c}{\textbf{correlate with attack effectiveness}}\\
\shortstack{\textbf{Factors $\rightarrow$}\\
\textbf{Attacks $\downarrow$}}  & 
\rotatebox{90}{\hyperlink{tr1:tail}{\textbf{\texttt{Tr1-Tail}}} \textbf{(Tail length)}} & 
\rotatebox{90}{\hyperlink{tr2:dataset}{\textbf{\texttt{Tr2-Data}}} \textbf{(Datasets)}} & 
\rotatebox{90}{\hyperlink{tr3:subgroups}{\textbf{\texttt{Tr3-Subgp}}} \textbf{(Subgroups)}} & 
\rotatebox{90}{\hyperlink{tr1:tail}{\textbf{\texttt{Tr1-Tail}}} \textbf{(Tail length)}} & 
\rotatebox{90}{\hyperlink{tr2:dataset}{\textbf{\texttt{Tr2-Data}}} \textbf{(Datasets)}} & 
\rotatebox{90}{\hyperlink{tr3:subgroups}{\textbf{\texttt{Tr3-Subgp}}} \textbf{(Subgroups)}}\\
% \rotatebox{90}{\hyperlink{req1:aux}{\textbf{\texttt{Te1-Aux}}} \textbf{(Aux. Data)}} & 
% \rotatebox{90}{\hyperlink{req2:dist}{\textbf{\texttt{Te2-Dist}}} \textbf{(Data Dstrb.)}} & 
% \rotatebox{90}{\hyperlink{req3:shadow}{\textbf{\texttt{Te3-Shdw}}} \textbf{(Model Fnct.)}} & 
% \rotatebox{90}{\hyperlink{req4:architecture}{\textbf{\texttt{Te4-Arch}}} \textbf{(Model Arch.)}} \\
\bottomrule

\toprule
% \multicolumn{8}{c}{\textbf{\underline{Train-time Risk}}} \\
\textbf{\poison} & $\uparrow$~\cite{yifeng2025swallowing} & $\uparrow$~\cite{tramer2022truth} & $\uparrow$~\cite{solans2020poisoning} & \cellcolor{gray!40} & \cellcolor{gray!40} & \cellcolor{gray!40} \\
\midrule
% \multicolumn{8}{c}{\textbf{\underline{Test-time Risks}}} \\
\textbf{\evasion} & \cellcolor{gray!40} & \cellcolor{gray!40} & \cellcolor{gray!40} & $\uparrow$~\cite{yue2024revisiting,cho2025longtailed} &  & \\
\midrule
\textbf{\mia} & \cellcolor{gray!40} & \cellcolor{gray!40} & \cellcolor{gray!40} & $\uparrow$~\cite{carlini2022membership,carlini2022the} & $\uparrow$~\cite{labelOnlyMIA,li2025enhanced} &  \\
\midrule
\textbf{\aia} & \cellcolor{gray!40} & \cellcolor{gray!40} & \cellcolor{gray!40} & $\uparrow$~\cite{jayaraman2022attribute} & & $\uparrow$~\cite{aalmoes2025alignment} \\
\midrule
\textbf{\dia} & \cellcolor{gray!40} &  \cellcolor{gray!40} & \cellcolor{gray!40} &  $\uparrow$~\cite{chase2021property,chaudhari2023snap} &  $\uparrow$~\cite{suri2022formalizing} & $\uparrow$ \cite{duddu2024sok} \\
\midrule
\textbf{\datarecon} & \cellcolor{gray!40}& \cellcolor{gray!40} & \cellcolor{gray!40} & $\uparrow$~\cite{li2025head} & $\uparrow$~\cite{ye2024leaveoneout} & $\uparrow$ \cite{duddu2024sok} \\
\midrule
\textbf{\modelext} & \cellcolor{gray!40} & \cellcolor{gray!40} & \cellcolor{gray!40} & $\downarrow$~\cite{same} & & \\
\midrule
\textbf{\disc} & \cellcolor{gray!40} & \cellcolor{gray!40} & \cellcolor{gray!40} & $\uparrow$~\cite{mehrabi2021survey} & & $\uparrow$~\cite{aalmoes2025alignment} \\
\bottomrule

\toprule
\end{tabular}
}
% \end{center}
\end{table}

\setlength{\tabcolsep}{2pt}
\begin{table}[ht]
\centering
\caption{\cosemtic{\textbf{(Test-time Collusion) Relation of attacks and factors:} \disc is excluded as there are no attacks that \adv can exploit. $\uparrow$ (or $\downarrow$) to indicate positive (negative) correlation; \colorbox{gray!20}{Gray} $\rightarrow$ not applicable; empty cells $\rightarrow$ no prior literature reported the correlations.}}
\label{tab:test_time_relation}
\vspace{0.25cm}
\resizebox{0.9\columnwidth}{!}{
\begin{tabular}{l||P{2cm}|P{2cm}|P{2cm}|P{2cm}}
\bottomrule

\toprule
& \multicolumn{4}{c}{\textbf{How attack effectiveness leads}} \\
& \multicolumn{4}{c}{\textbf{to change in factors (as ``outcomes'')}}\\ 
% & \multicolumn{4}{c}{\textbf{Correlation of attack effectiveness}} \\
% & \multicolumn{4}{c}{\textbf{with change in factors}} \\
\shortstack{\textbf{Factors $\rightarrow$}\\
\textbf{Attacks $\downarrow$}} & 
\rotatebox{90}{\hyperlink{req1:aux}{\textbf{\texttt{Te1-Aux}}} \textbf{(Aux. Data)}} & 
\rotatebox{90}{\hyperlink{req2:dist}{\textbf{\texttt{Te2-Dist}}} \textbf{(Data Dstrb.)}} & 
\rotatebox{90}{\hyperlink{req3:shadow}{\textbf{\texttt{Te3-Shdw}}} \textbf{(Model Fnct.)}} & 
\rotatebox{90}{\hyperlink{req4:architecture}{\textbf{\texttt{Te4-Arch}}} \textbf{(Model Arch.)}} \\
\bottomrule

\toprule
\textbf{\evasion} & $\uparrow$~\cite{prada,labelOnlyMIA} &  & & \\
\midrule
\textbf{\mia} & $\uparrow$~\cite{shokri2017membership,mlleaks} &  & & \\
\midrule
\textbf{\aia} &  & $\uparrow$~\cite{aalmoes2025alignment}  & & \\
\midrule
\textbf{\dia} &   & $\uparrow$~\cite{suri2022formalizing,suri2023dissecting} &  & \\
\midrule
\textbf{\datarecon} & $\uparrow$~\cite{sokDataRecon,sokGradientRecon} & & & \\
\midrule
\textbf{\modelext} &  & & $\uparrow$~\cite{orekondy2019knockoff,prada} & $\uparrow$~\cite{orekondy2019knockoff,prada} \\
\bottomrule

\toprule
\textbf{Factors $\rightarrow$} & \multicolumn{4}{c}{\textbf{How change in factors (as ``prerequisites'')}} \\
\textbf{Attacks $\downarrow$} & \multicolumn{4}{c}{\textbf{correlate with attack effectiveness}}\\\\
% & \multicolumn{4}{c}{\textbf{Correlation of change in factors}} \\
% & \multicolumn{4}{c}{\textbf{with attack effectiveness}} \\\\
% \textbf{Attack} & 
% \rotatebox{90}{\hyperlink{req1:aux}{\textbf{\texttt{Te1-Aux}}} \textbf{(Aux. Data)}} & 
% \rotatebox{90}{\hyperlink{req2:dist}{\textbf{\texttt{Te2-Dist}}} \textbf{(Data Dstrb.)}} & 
% \rotatebox{90}{\hyperlink{req3:shadow}{\textbf{\texttt{Te3-Shdw}}} \textbf{(Model Fnct.)}} & 
% \rotatebox{90}{\hyperlink{req4:architecture}{\textbf{\texttt{Te4-Arch}}} \textbf{(Model Arch.)}} \\
% \bottomrule
% \toprule
\textbf{\evasion} & $\uparrow$~\cite{demontis2019why,pmlrwu20a,gu2024a} & $\uparrow$~\cite{pmlrwu20a,demontis2019why,gu2024a} & $\uparrow$~\cite{pmlrwu20a,demontis2019why,gu2024a} & $\uparrow$~\cite{pmlrwu20a,demontis2019why,gu2024a} \\
\midrule
\textbf{\mia} & $\uparrow$~\cite{shokri2017membership,mlleaks} & $\uparrow$~\cite{shokri2017membership,mlleaks} & $\uparrow$~\cite{shokri2017membership,mlleaks} & $\uparrow$~\cite{shokri2017membership,mlleaks} \\
\midrule
\textbf{\aia} & $\uparrow$~\cite{jayaraman2022attribute,liu2022ml} &  $\uparrow$~\cite{jayaraman2022attribute,liu2022ml}  & $\uparrow$~\cite{liu2022ml} & $\uparrow$~\cite{liu2022ml} \\
\midrule
\textbf{\dia} & $\uparrow$~\cite{suri2022formalizing,suri2023dissecting} &  & $\uparrow$~\cite{suri2022formalizing,suri2023dissecting} & $\uparrow$~\cite{suri2022formalizing,suri2023dissecting} \\
\midrule
\textbf{\datarecon} & $\uparrow$~\cite{sokDataRecon,sokGradientRecon} & $\uparrow$~\cite{sokDataRecon,sokGradientRecon} & $\uparrow$~\cite{sokDataRecon,sokGradientRecon} & $\uparrow$~\cite{sokDataRecon,sokGradientRecon} \\
\midrule
\textbf{\modelext} & $\uparrow$~\cite{orekondy2019knockoff,prada} & $\uparrow$~\cite{orekondy2019knockoff,prada} &   &  \\
\bottomrule

\toprule
\end{tabular}
}
\vspace{-0.5cm}
\end{table}

%% file: 4systematization.tex
\section{Systematization of Prior Work}\label{sec:systematization}

We apply our guideline to explain collusion in prior work using factors from our framework (\Snospace\ref{sec:traintestcollusion} and \Snospace\ref{sec:testcollusion}).
We identify prior work through a literature survey and describe our methodology for collecting papers below.

\noindent\emph{Methodology for Collecting Papers:} We surveyed papers covering collusion among adversaries. 
We collected an initial set of papers using keywords containing both risks. For instance, ``poisoning/backdoors increases <\attacktwo>'' in case of train-test collusions, and ``\attackone increases \attacktwo'' for test-time collusions.
We then reviewed the citations of these initial papers for comprehensive coverage. Overall, we collected 28 and 12 papers for train-test and test-time collusion respectively.

\subsection{Train-Test Collusions}\label{sec:traintestcollusion}

% Table~\ref{tab:train_test_interactions} shows various possible train-test collusions.
% We also indicate the common enabling factors identified using our guideline to explain the potential for collusion, i.e., exploiting \texttt{Risk1} increases \texttt{Risk2} (indicated as $\uparrow$, $\uparrow$).
Table~\ref{tab:train_test_interactions} shows train-test collusion (indicated as ``\attackone $\rightarrow$ \attacktwo'' where \attackone is always \poison).
We indicate the common enabling factors identified using the guideline to explain the collusion indicated as ($\uparrow$, $\uparrow$) corresponding to (<$\cdot$>$_1$, <$\cdot$>$_2$) where <$\cdot$>$_1$ corresponds to the arrow depicting the ``How attack effectiveness leads to change in factors'' (for \attackone) while <$\cdot$>$_2$ corresponds to the arrow depicting the ``How changes in factors correlate with attack effectiveness'' (for \attacktwo).

\input{tables/tab_train_test_interactions}
\begin{itemize}[leftmargin=*]
    \item \textbf{\poison/\backdoor}$\rightarrow$\textbf{\evasion}: \poison increases the susceptibility to \evasion by increasing $\dtrain$'s tail length (\hyperlink{tr1:tail}{\textbf{\texttt{Tr1-Tail}}}: $\uparrow$, $\uparrow$)~\cite{jiang2024adv,pang2020tale,fowl2021adversarial}.
    
    \item \textbf{\poison/\backdoor}$\rightarrow$\textbf{\mia}: \poison increases susceptibility to \mia by increasing $\dtrain$'s tail length, and distinguishability between data records inside and outside $\dtrain$ (\hyperlink{tr1:tail}{\textbf{\texttt{Tr1-Tail}}}, \hyperlink{tr2:dataset}{\textbf{\texttt{Tr2-Data}}}: $\uparrow$, $\uparrow$)~\cite{tramer2022truth,chen2025codepoisonMIA,wen2024privacy,chen2022amplifying,zhang2023agrevader}.
    
    \item \textbf{\poison}$\rightarrow$\textbf{\aia}: \poison increases susceptibility to \aia by increasing $\dtrain$'s tail length, and distinguishability across demographic subgroups (\hyperlink{tr1:tail}{\textbf{\texttt{Tr1-Tail}}}, \hyperlink{tr3:subgroups}{\textbf{\texttt{Tr3-Subgp}}}: $\uparrow$, $\uparrow$)~\cite{tramer2022truth,malekzadeh2021honest,ding2023vertexserum}.
    
    \item \textbf{\poison}$\rightarrow$\textbf{\dia}: \poison increases susceptibility to \dia by increasing $\dtrain$'s tail length, and distinguishability across datasets with different distributions (\hyperlink{tr1:tail}{\textbf{\texttt{Tr1-Tail}}}, \hyperlink{tr2:dataset}{\textbf{\texttt{Tr2-Data}}}: $\uparrow$, $\uparrow$)~\cite{poisoningPropInfFL,chase2021property,chaudhari2023snap,tian2023cvpr,luo2024exploring}.
    
    \item \textbf{\poison/\backdoor}$\rightarrow$\textbf{\datarecon}: \poison/\backdoor make \datarecon more effective by increasing $\dtrain$'s tail length, and increasing the distinguishability across datasets and subgroups (\hyperlink{tr1:tail}{\textbf{\texttt{Tr1-Tail}}}, \hyperlink{tr2:dataset}{\textbf{\texttt{Tr2-Data}}}, \hyperlink{tr3:subgroups}{\textbf{\texttt{Tr3-Subgp}}}: $\uparrow$, $\uparrow$)~\cite{tramer2022truth,panda2024teach,feng2024privacy,peng2024data,loki,song2017machine,küchler2025architectural}.
    
    \item \textbf{\poison/\backdoor}$\rightarrow$\textbf{\disc}: \poison/\backdoor increases \disc by increasing in $\dtrain$'s tail length, and distinguishability across different subgroups (\hyperlink{tr1:tail}{\textbf{\texttt{Tr1-Tail}}},\hyperlink{tr3:subgroups}{\textbf{\texttt{Tr3-Subgp}}}: $\uparrow$, $\uparrow$)~\cite{naseh2024backdooring,mehrabi2021exacerbating,solans2020poisoning,ijcai2024p51,furth,zheng2023trojfair,wang2025}.
\end{itemize}

\input{tables/tab_test_interactions}

\subsection{Test-Time Collusions}\label{sec:testcollusion}

% Table~\ref{tab:test_interactions} shows test-time collusions and common factors for test-time risks: \texttt{Risk1} and \texttt{Risk2}.
% Our guideline explains the potential for collusion in prior work by identifying prerequisites of \texttt{Risk2} which are satisfied by the outcomes from exploiting \texttt{Risk1} (indicated as $\uparrow$, $\uparrow$).
Table~\ref{tab:test_interactions} shows test-time collusion (indicated as ``\attackone $\rightarrow$ \attacktwo''). We denote the common factors as (<$\cdot$>$_1$, <$\cdot$>$_2$) where <$\cdot$>$_1$ corresponds to the arrow depicting the ``How attack effectiveness leads to change in factors'' (for \attackone) while <$\cdot$>$_2$ corresponds to the arrow depicting the ``How change
in factors correlate with attack effectiveness'' (for \attacktwo).
% The guideline explains collusion in prior work by identifying the prerequisites for \attacktwo which are satisfied by the outcomes of \attackone, indicated as ($\uparrow$, $\uparrow$).}
\begin{itemize}[leftmargin=*]
    \item \textbf{\evasion} $\rightarrow$ \{\textbf{\text{\modelext}}, \textbf{\text{\mia}}, \textbf{\text{\dia}}\}: 
    \modelext, \mia, and \dia require $\daux$ to train $\shadowmodel$ (\hyperlink{req1:aux}{\textbf{\texttt{Te1-Aux}}}: $\uparrow$). 
    \evasion can provide $\daux$ with adversarial examples to meet the prerequisites of \modelext, \mia, and \dia (\hyperlink{req1:aux}{\textbf{\texttt{Te1-Aux}}}: $\uparrow$)~\cite{prada,liu2025amplifying}.

    \item \textbf{\mia} $\rightarrow$ \{\text{\textbf{\aia}}, \text{\textbf{\datarecon}}, \text{\textbf{\modelext}}\}:
    \aia,  \datarecon, and \text{\textbf{\modelext}} can be more effective when $\daux$ has some overlap with $\dtrain$ to increase the functional similarity of $\shadowmodel$ with $\model$ (\hyperlink{req1:aux}{\textbf{\texttt{Te1-Aux}}}: $\uparrow$). 
    \adv{1} can exploit \mia to identify data records in $\dtrain$ which can be added to $\daux$ to improve the functional similarity of $\shadowmodel$ to $\model$. This can help design more effective (and transferrable) attacks: \aia, \datarecon, and \text{\textbf{\modelext}} (\hyperlink{req1:aux}{\textbf{\texttt{Te1-Aux}}}: $\uparrow$)~\cite{yeom2018privacy, extractlm, extractionDiffusion}. 
    % Additionally, \aia can be implemented by using \mia as a subroutine~\cite{aiaMIa}. Hence, \mia can enable the effectiveness of \aia.
    
    \item \textbf{\dia} $\rightarrow$ \{\text{\textbf{\aia}}, \text{\textbf{\mia}}\}:
    \aia and \mia are more effective if $\daux$ has a similar distribution as $\dtrain$ (\hyperlink{req2:dist}{\textbf{\texttt{Te2-Dist}}}: $\uparrow$). 
    By exploiting \dia, \adv{1} can infer $\dtrain$'s distribution, and \adv{2} use that knowledge to update $\daux$ to train $\shadowmodel$ which is functionally more similar to $\model$. This can improve the effectiveness of \aia and \mia (\hyperlink{req2:dist}{\textbf{\texttt{Te2-Dist}}}: $\uparrow$)~\cite{liu2025amplifying}.
    
    \item \textbf{\datarecon} $\rightarrow$ \text{\textbf{\modelext}}: 
    Training $\shadowmodel$ on $\daux$ which partially overlaps with $\dtrain$, increases \modelext's effectiveness (\hyperlink{req1:aux}{\textbf{\texttt{Te1-Aux}}}: $\uparrow$).
    \datarecon can identify data records in $\dtrain$ which can be added to $\daux$. This increases the overlap with $\dtrain$ improving the functional similarity of $\shadowmodel$ with $\model$, enabling a more effective \modelext (\hyperlink{req1:aux}{\textbf{\texttt{Te1-Aux}}}: $\uparrow$)~\cite{gong2021inversenet}.

    \item \textbf{\modelext} $\rightarrow$ \{\text{\textbf{\evasion}}, \text{\textbf{\mia}}, \text{\textbf{\aia}}\}: 
    Transferable adversarial examples (in a blackbox setting), \mia, \aia, and \datarecon require $\shadowmodel$ which is functionally similar to $\text{$\model$}$ (\hyperlink{req3:shadow}{\textbf{\texttt{Te3-Shdw}}}, \hyperlink{req4:architecture}{\textbf{\texttt{Te4-Arch}}}: $\uparrow$).
    \modelext can help extract such $\shadowmodel$ from $\model$ to meet the prerequisites of other attacks (\hyperlink{req3:shadow}{\textbf{\texttt{Te3-Shdw}}}, \hyperlink{req4:architecture}{\textbf{\texttt{Te4-Arch}}}: $\uparrow$)~\cite{papernot2017practical, prada, xiao2022mexmi, chen2021killing}.
\end{itemize}

%% file: tables/tab_train_test_interactions.tex
\setlength{\tabcolsep}{4pt}
\begin{table}[ht!]
\caption{\textbf{Train-Test Collusions:} Collusion between \adv{1} (exploits \poison) and \adv{2} (exploits test-time \attacktwo). \textbf{``Factors''} $\rightarrow$ common enabling factors (Tables~\ref{tab:train_test_relation} and~\ref{tab:test_time_relation}) to identify potential for collusion ($\uparrow$, $\uparrow$).
\colorbox{gray!20}{Gray} $\rightarrow$ no prior work.}
\vspace{0.25cm}
\centering
% \begin{center}
\begin{scriptsize}
\resizebox{\columnwidth}{!}{
\begin{tabular}{l|l|c|c}
\bottomrule

\toprule
\textbf{\attacktwo} & \textbf{References} & \textbf{Factors} (<\attackone>, <\attacktwo>) & \textbf{Collusion?}\\
\bottomrule

\toprule

\multirow{2}{*}{\textbf{\evasion}} & \multirow{2}{*}{\cite{jiang2024adv,pang2020tale,fowl2021adversarial}} & 
\multirow{2}{*}{\shortstack{\hyperlink{tr1:tail}{\textbf{\texttt{Tr1-Tail}}}\\($\uparrow$, $\uparrow$)}} & \multirow{2}{*}{Yes}\\
& & \\

\midrule
\cellcolor{gray!20}\textbf{\modelext} & \cellcolor{gray!20}\textbf{Our Work (\Snospace\ref{sec:trteeval})} & 
\cellcolor{gray!20}\hyperlink{tr1:tail}{\textbf{\texttt{Tr1-Tail}}} ($\uparrow$, $\downarrow$) & No\\

\midrule
\multirow{2}{*}{\textbf{\mia}} & \multirow{2}{*}{\cite{tramer2022truth,chen2025codepoisonMIA,wen2024privacy,chen2022amplifying,zhang2023agrevader}} & 
\multirow{2}{*}{\shortstack{\hyperlink{tr1:tail}{\textbf{\texttt{Tr1-Tail}}}, \hyperlink{tr2:dataset}{\textbf{\texttt{Tr2-Data}}}\\($\uparrow$, $\uparrow$)}}  & \multirow{2}{*}{Yes}\\
& & \\

\midrule
\multirow{2}{*}{\textbf{\aia}} & \multirow{2}{*}{\cite{tramer2022truth,malekzadeh2021honest,ding2023vertexserum}} & 
\multirow{2}{*}{\shortstack{\hyperlink{tr1:tail}{\textbf{\texttt{Tr1-Tail}}}, \hyperlink{tr3:subgroups}{\textbf{\texttt{Tr3-Subgp}}}\\($\uparrow$, $\uparrow$)}}  & \multirow{2}{*}{Yes}\\
& & \\

\midrule
\multirow{2}{*}{\textbf{\dia}} & \multirow{2}{*}{\cite{poisoningPropInfFL,chase2021property,chaudhari2023snap,tian2023cvpr,luo2024exploring}} & 
\multirow{2}{*}{\shortstack{\hyperlink{tr1:tail}{\textbf{\texttt{Tr1-Tail}}}, \hyperlink{tr3:subgroups}{\textbf{\texttt{Tr3-Subgp}}}\\($\uparrow$, $\uparrow$)}}  & \multirow{2}{*}{Yes}\\
& & \\

\midrule
\multirow{2}{*}{\textbf{\datarecon}} & \multirow{2}{*}{\cite{tramer2022truth,panda2024teach,feng2024privacy,peng2024data,loki,song2017machine,küchler2025architectural}} & 
\multirow{2}{*}{\shortstack{\hyperlink{tr1:tail}{\textbf{\texttt{Tr1-Tail}}}, \hyperlink{tr2:dataset}{\textbf{\texttt{Tr2-Data}}}, \\ \hyperlink{tr3:subgroups}{\textbf{\texttt{Tr3-Subgp}}} ($\uparrow$, $\uparrow$)}}  & \multirow{2}{*}{Yes}\\
& & \\

\midrule
\multirow{2}{*}{\textbf{\disc}} & \multirow{2}{*}{\cite{naseh2024backdooring,mehrabi2021exacerbating,solans2020poisoning,ijcai2024p51,furth,zheng2023trojfair,wang2025}} & 
\multirow{2}{*}{\shortstack{\hyperlink{tr1:tail}{\textbf{\texttt{Tr1-Tail}}}, \hyperlink{tr3:subgroups}{\textbf{\texttt{Tr3-Subgp}}}\\($\uparrow$, $\uparrow$)}}  & \multirow{2}{*}{Yes} \\
& & \\

\bottomrule

\toprule
\end{tabular}
}
\end{scriptsize}
% \end{center}
\vspace{-0.5cm}
\label{tab:train_test_interactions}
\end{table}

%% file: tables/tab_test_interactions.tex
\setlength{\tabcolsep}{2pt}
\begin{table}[!htb]
\caption{\textbf{Test-time Collusions:} Both adversaries exploit test-time risks: \attackone and \attacktwo. \textbf{``Factors''} $\rightarrow$ common test-time enabling factors to see potential for collusion (prerequisite of \attacktwo is satisfied by \attackone: $\uparrow$, $\uparrow$).
\colorbox{gray!20}{Gray} $\rightarrow$ no prior work.
% \colorbox{red!20}{NC} $\rightarrow$ no common factor.
}
\vspace{0.25cm}
\centering
% \begin{center}
% \def\arraystretch{0.9}
\begin{scriptsize}
\begin{tabular}{l|l|P{2.5cm}|c}
\bottomrule

\toprule
\attackone & \attacktwo & \textbf{Factors} (<\texttt{Risk1}>, <\texttt{Risk2}>) & \textbf{Collusion?}\\
\bottomrule

\toprule

\multirow{5}{*}{\textbf{\evasion}} & \textbf{\modelext}~\cite{prada} & \multirow{5}{*}{\shortstack{\hyperlink{req1:aux}{\textbf{\texttt{Te1-Aux}}}\\($\uparrow$, $\uparrow$)}} & \multirow{5}{*}{Yes}\\
  & \textbf{\mia}~\cite{liu2025amplifying} & \\
 & \cellcolor{gray!20}\textbf{\aia}  & \\
 & \textbf{\dia}~\cite{liu2025amplifying} & \\
  & \cellcolor{gray!20}\textbf{\datarecon}  & \\
 % & \cellcolor{gray!20}\textbf{\disc}  & \cellcolor{red!20}NC \\
\midrule

 \multirow{5}{*}{\textbf{\modelext}} & \textbf{\evasion}~\cite{prada,papernot2017practical} & 
\multirow{5}{*}{\shortstack{\hyperlink{req3:shadow}{\textbf{\texttt{Te3-Shdw}}}, \hyperlink{req4:architecture}{\textbf{\texttt{Te4-Arch}}}\\($\uparrow$, $\uparrow$)}} & \multirow{5}{*}{Yes}\\
 & \textbf{\mia}~\cite{xiao2022mexmi} & \\
& \textbf{\aia}~\cite{chen2021killing} & \\
 & \textbf{\dia}\cellcolor{gray!20} & \\
 & \textbf{\datarecon}\cellcolor{gray!20} & \\
 % & \textbf{\disc}\cellcolor{gray!20} & \cellcolor{red!20}NC  \\ 
\midrule

 \multirow{5}{*}{\textbf{\mia}} & \textbf{\evasion} & \multirow{5}{*}{\shortstack{\hyperlink{req1:aux}{\textbf{\texttt{Te1-Aux}}}\\($\uparrow$, $\uparrow$)}} & \multirow{5}{*}{Yes} \\ 
& \textbf{\modelext}~\cite{xiao2022mexmi} & \\ 
 & \textbf{\aia}~\cite{yeom2018privacy,aiaMIa} & \\
 & \cellcolor{gray!20}\textbf{\dia} & \\ 
& \textbf{\datarecon}~\cite{extractlm,extractionDiffusion} & \\
 % & \textbf{\disc}\cellcolor{gray!20}  & \cellcolor{red!20}NC\\ 
\midrule

\multirow{5}{*}{\textbf{\aia}} & \cellcolor{gray!20}\textbf{\evasion} & \multirow{5}{*}{\shortstack{\hyperlink{req2:dist}{\textbf{\texttt{Te2-Dist}}}\\($\uparrow$, $\uparrow$)}} & \multirow{5}{*}{Yes}\\
& \cellcolor{gray!20}\textbf{\modelext}  & \\
 & \cellcolor{gray!20}\textbf{\mia} & \\ 
& \cellcolor{gray!20}\textbf{\dia}  & \\
 & \cellcolor{gray!20}\textbf{\datarecon} & \\
% & \cellcolor{gray!20}\textbf{\disc} & \cellcolor{red!20}NC \\
\midrule

 \multirow{5}{*}{\textbf{\dia}} & \cellcolor{gray!20}\textbf{\evasion}  & \multirow{5}{*}{\shortstack{\hyperlink{req2:dist}{\textbf{\texttt{Te2-Dist}}}\\($\uparrow$, $\uparrow$)}} & \multirow{5}{*}{Yes}\\
 & \cellcolor{gray!20}\textbf{\modelext}  & \\
& \textbf{\mia}~\cite{liu2025amplifying} & \\
& \textbf{\aia}~\cite{liu2025amplifying} & \\
& \cellcolor{gray!20}\textbf{\datarecon} & \\
 % & \cellcolor{gray!20}\textbf{\disc}  & \cellcolor{red!20}NC \\
\midrule

 \multirow{5}{*}{\textbf{\datarecon}} & \cellcolor{gray!20}\textbf{\evasion} & \multirow{5}{*}{\shortstack{\hyperlink{req1:aux}{\textbf{\texttt{Te1-Aux}}}\\($\uparrow$, $\uparrow$)}} & \multirow{5}{*}{Yes}\\
& \textbf{\modelext}~
\cite{gong2021inversenet} & \\
 & \textbf{\mia}\cellcolor{gray!20} & \\
 & \textbf{\aia}\cellcolor{gray!20}  & \\
 & \textbf{\dia}\cellcolor{gray!20}  & \\
 % & \textbf{\disc}\cellcolor{gray!20} & \cellcolor{red!20}NC \\
% \midrule

% \multirow{5}{*}{\textbf{\disc}} & \cellcolor{gray!20}\textbf{\evasion} & \multirow{5}{*}{\shortstack{\hyperlink{req2:dist}{\textbf{\texttt{Te2-Dist}}}\\($\uparrow$, $\uparrow$)}} \\ 
%  & \cellcolor{gray!20}\textbf{\modelext}  & \\
%  & \cellcolor{gray!20}\textbf{\mia}  & \\
%  & \textbf{\aia}~\cite{aalmoes2025alignment} & \\
%  & \textbf{\datarecon}~\cite{duddu2024sok} & \\
%  & \textbf{\dia}  & \textbf{Our Work (\Snospace\ref{sec:teeval})}\\
\bottomrule

\toprule
\end{tabular}
\end{scriptsize}
% \end{center}
\vspace{-0.75cm}
\label{tab:test_interactions}
\end{table}

%% file: 5evaluation.tex
\section{Unexplored Collusions among Adversaries}\label{sec:evaluation}

We apply our guideline to conjecture about the potential of unexplored collusions (\Snospace\ref{sec:unexplored} and marked as \colorbox{gray!25}{gray} in Tables~\ref{tab:train_test_interactions} and~\ref{tab:test_interactions}), and empirically validate one train-test and \change{four} test-time collusions (\Snospace\ref{sec:trteeval}, \Snospace\ref{sec:teeval2}, and \Snospace\ref{sec:teeval3})\footnote{Code: \url{https://github.com/ssg-research/sok-collusion}}.

\subsection{Conjectures for Unexplored Collusions}\label{sec:unexplored}

\noindent\textbf{\underline{Train-Test Collusions:}} There is only one unexplored train-test collusion (\colorbox{gray!25}{gray} in Table~\ref{tab:train_test_interactions}) between \adv{1} who executes \textbf{\poison}, and \adv{2} who executes \textbf{\modelext}. 
We see that \hyperlink{tr1:tail}{\textbf{\texttt{Tr1-Tail}}} is the common factor between both attacks.
Since the arrows are opposing ($\uparrow$ for \poison and $\downarrow$ for \modelext), our guideline suggests that there is no potential for collusion (\poison decreases the effectiveness of \modelext).
We validate this conjecture in \Snospace\ref{sec:trteeval}.

\noindent\textbf{\underline{Test-time Collusions:}} We discuss various unexplored test-time collusions (\colorbox{gray!25}{gray} in Table~\ref{tab:test_interactions}), and use our guideline to identify the potential for collusion, indicated as ($\uparrow$, $\uparrow$).
\begin{itemize}[leftmargin=*]
    \item \textbf{\evasion} $\rightarrow$ \{\textbf{\aia}, \textbf{\datarecon}\}: \hyperlink{req1:aux}{\textbf{\texttt{Te1-Aux}}} is the common factor which suggests that \evasion generates adversarial examples which can be used as $\daux$ to train $\shadowmodel$. Since the adversarial examples can characterize $\model$'s decision boundary, it can allow $\shadowmodel$ to be functionally similar to $\model$. The resulting $\shadowmodel$ can be used as a ``shadow model'' for other attacks such as \aia, \datarecon, and \disc.

    \item \textbf{\mia} $\rightarrow$ \{\textbf{\evasion}, \textbf{\dia}\}: \mia can help identify data records in $\dtrain$, which can then be included as part of $\daux$ (\hyperlink{req1:aux}{\textbf{\texttt{Te1-Aux}}}). 
    The resulting $\shadowmodel$ trained on $\daux$ which overlaps with $\dtrain$, is likely to increase the functional similarity with $\model$, thereby improving the effectiveness of \evasion, \modelext, and \dia.For instance, prior work on \evasion and \dia have shown that increasing overlap between $\daux$ and $\dtrain$ can increase the effectiveness of the attacks~\cite{demontis2019why,suri2022formalizing,leakage21Prop}.

    \item \textbf{\aia} $\rightarrow$ \{\textbf{\evasion}, \textbf{\mia}, \textbf{\dia}, \textbf{\datarecon}, \textbf{\modelext}\}: \aia can help estimate distribution of $\dtrain$ allowing \adv to subsequently update $\daux$ (\hyperlink{req2:dist}{\textbf{\texttt{Te2-Dist}}}). This can improve the effectiveness of \evasion, \modelext, \mia, \datarecon, and \dia, by training $\shadowmodel$ which is functionally more similar to $\model$.

    \item \textbf{\dia} $\rightarrow$ \{\textbf{\evasion}, \textbf{\datarecon}, \textbf{\modelext}\}: \dia can infer $\dtrain$'s distribution allowing \adv to subsequently update $\daux$ (\hyperlink{req2:dist}{\textbf{\texttt{Te2-Dist}}}). 
    The resulting $\shadowmodel$ is likely to be more similar to $\model$, thus, likely improving the effectiveness of \evasion, \modelext, \datarecon, and \dia.

    \item \textbf{\datarecon} $\rightarrow$ \{\textbf{\evasion}, \textbf{\mia}, \textbf{\aia}, \textbf{\dia}\}: \hyperlink{req1:aux}{\textbf{\texttt{Te1-Aux}}} is the common factor among the attacks. 
    \datarecon can help partially identify data records in $\dtrain$ which can augment $\daux$ (\hyperlink{req1:aux}{\textbf{\texttt{Te1-Aux}}}). 
    This can result in training $\shadowmodel$ to be similar to $\model$, thus, improving the effectiveness of \evasion, \mia, \aia, and \dia.

    \item \textbf{\modelext} $\rightarrow$ \{\textbf{\dia}, \textbf{\datarecon}\}: \modelext can help train $\shadowmodel$ that is similar to $\model$ (\hyperlink{req3:shadow}{\textbf{\texttt{Te3-Shdw}}}, \hyperlink{req4:architecture}{\textbf{\texttt{Te4-Arch}}}), which can be used as ``shadow models'' for effective \dia and \datarecon.
    
\end{itemize}
As an illustrative example, we validate the conjecture for \change{four test-time collusions: \modelext $\rightarrow$ \dia (\Snospace\ref{sec:teeval2}) and \datarecon $\rightarrow$ \{\mia, \aia, \dia\} (\Snospace\ref{sec:teeval3}).}

\begin{takeaway}
    \textbf{Takeaway:} Unexplored collusions are gaps in the current literature, which are left as directions for future work.
\end{takeaway}

\subsection{Train-Test Collusion: \texorpdfstring{\poison}{Poison} \texorpdfstring{$\rightarrow$}{right arrow }\texorpdfstring{\modelext}{ModExt}}\label{sec:trteeval}

We now validate our conjecture that there is no potential for collusion between \poison and \modelext, marked as ($\uparrow$, $\downarrow$) in Table~\ref{tab:train_test_interactions}.
Since \poison increases the tail length of $\dtrain$ (\hyperlink{tr1:tail}{\textbf{\texttt{Tr1-Tail}}}), we conjecture that $\surrogatemodel$ may fail to learn the under-represented data records, thereby reducing the effectiveness of \modelext.

\noindent\textbf{\underline{Experimental Setup:}}
We run our experiments on two image datasets: CIFAR10/100. The dataset contains 60,000 images (50,000 for training and 10,000 for testing) where each data point is a 32~$\times$~32 colored image. The images are clustered into 10 and 100 classes for CIFAR10 and CIFAR100 respectively, representing different objects. 
We use 25,000 records for $\dtrain$ and the remaining 25,000 records as $\daux$ for training $\surrogatemodel$ via \modelext. 
Hence, the accuracy of $\model$ is not the same as the state-of-the-art on these datasets.
We use ResNet models as $\model$ and VGG16 as $\surrogatemodel$.
We check that $\model$ has been successfully poisoned using \emph{accuracy on poisoned test dataset}\footnote{We empirically verify this, but do not report the numbers for brevity.}. We report $\surrogatemodel$'s \emph{accuracy} and \emph{fidelity} (the percentage of outputs that match between $\model$ and $\surrogatemodel$) on a clean test dataset.
We use backdoors with a 5 x 5 white patch as the trigger, and knockoff-nets for \modelext by training $\surrogatemodel$ using $\model$'s predictions as the ground truth~\cite{orekondy2019knockoff}. We consider different query budgets for training $\surrogatemodel$ including \{10\%, 25\%, 50\%, 100\%\} of the 25,000 data records in $\daux$.

\input{tables/tab_trte_eval}

\noindent\textbf{\underline{Results:}}
Table~\ref{tab:trteeval} shows the results of extracting the target model $\model$ using \modelext on CIFAR10/100.
For a given budget, we compare the surrogate accuracy (``Surr. Acc.'') and fidelity (``Surr. Fid.'') when $\model$ is trained with poisons \{0.05\%, 0.1\%, 0.15\%, 0.2\%\}, with the case where $\model$ is trained without poisons (0\% poison). 
We select a low poison rate to evaluate \modelext's effect, without significant utility degradation from poisoning.
For a specific budget, to compare the baseline accuracy and fidelity (i.e., 0\% poison), we use \colorbox{green!15}{green} for higher than baseline, \colorbox{orange!15}{orange} for similar (within standard deviation), and \colorbox{red!15}{red} for lower than baseline.

Across different poison rates and budgets, we see that the surrogate accuracy and fidelity are lower than the baseline for CIFAR100 (\colorbox{red!15}{red}).
For CIFAR10, the surrogate accuracy and fidelity are similar to the baseline (\colorbox{orange!15}{orange}) for the low budget of 2500 queries, where $\surrogatemodel$ generalizes well. However, for higher budgets, the accuracy / fidelity are lower than the baseline (\colorbox{red!15}{red}).
Overall, these results suggest that \poison reduces the effectiveness of \modelext, confirming our conjecture that there is no increase in potential for collusion.

\input{new_experiments}

%% file: tables/tab_trte_eval.tex
\begin{table}[!htb]
\caption{\textbf{\poison} $\rightarrow$ \textbf{\modelext}: Surrogate model's accuracy (``Surr. Acc.'') and fidelity (``Surr. Fid.'') for 0\% poison are the baseline. For a given budget, \colorbox{green!15}{green} $\rightarrow$ higher than baseline, \colorbox{orange!15}{orange} $\rightarrow$ similar to baseline (within std. dev.), and \colorbox{red!15}{red} $\rightarrow$ lower than baseline. We report target accuracy to show that \modelext is effective for different poisoning rates.}
\begin{center}
\resizebox{0.8\columnwidth}{!}{
\begin{tabular}{l|c|c|c|c}
\bottomrule

\toprule
& \multicolumn{2}{c|}{\textbf{CIFAR10}} & \multicolumn{2}{c}{\textbf{CIFAR100}}\\
\textbf{Budget} &
\textbf{Surr. Acc.} & \textbf{Surr. Fid.} &  \textbf{Surr. Acc.} & \textbf{Surr. Fid.}\\
\bottomrule

\toprule
\multicolumn{5}{c}{\textbf{0\% Poison (Baseline)}}\\
& \multicolumn{2}{c|}{\textbf{Target Acc.}: $87.37_{\pm 0.16}$} & \multicolumn{2}{c}{\textbf{Target Acc.}: $62.24_{\pm 0.47}$}\\
\midrule
2500 & $76.83_{\pm 0.71}$ & $78.54_{\pm 0.54}$ & $46.99_{\pm 1.59}$ & $54.75_{\pm 1.16}$\\
6250 & $83.41_{\pm 0.23}$ & $87.19_{\pm 1.09}$ & $53.04_{\pm 1.07}$ & $62.63_{\pm 0.92}$\\
12500 & $85.05_{\pm 0.27}$ & $88.82_{\pm 0.63}$ & $54.67_{\pm 0.84}$ & $65.40_{\pm 0.70}$\\
25000 & $85.91_{\pm 0.43}$ & $89.58_{\pm 1.07}$ & $56.09_{\pm 0.88}$ & $66.55_{\pm 0.90}$\\
\midrule
\multicolumn{5}{c}{\textbf{5\% Poison}}\\
& \multicolumn{2}{c|}{\textbf{Target Acc.}: $78.19_{\pm 0.60}$} & \multicolumn{2}{c}{\textbf{Target Acc.}: $33.69_{\pm 1.36}$}\\
\midrule
2500  & \cellcolor{orange!15}$75.30_{\pm 1.12}$ & \cellcolor{orange!15}$79.84_{\pm 1.71}$ & \cellcolor{red!15}$33.25_{\pm 0.29}$ & \cellcolor{red!15}$37.24_{\pm 0.60}$\\
 6250 & \cellcolor{red!15}$78.04_{\pm 0.42}$ & \cellcolor{red!15}$83.31_{\pm 0.85}$ &  \cellcolor{red!15}$35.00_{\pm 0.13}$ & \cellcolor{red!15}$39.90_{\pm 0.59}$\\
 12500 & \cellcolor{red!15}$78.30_{\pm 0.25}$ & \cellcolor{red!15}$83.31_{\pm 0.31}$ &  \cellcolor{red!15}$35.34_{\pm 1.21}$ & \cellcolor{red!15}$40.81_{\pm 1.22}$\\
25000 & \cellcolor{red!15}$78.93_{\pm 0.26}$ & \cellcolor{red!15}$84.39_{\pm 0.24}$ & \cellcolor{red!15}$34.80_{\pm 0.55}$ & \cellcolor{red!15}$40.14_{\pm 1.41}$\\
\midrule

\multicolumn{5}{c}{\textbf{10\% Poison}}\\
& \multicolumn{2}{c|}{\textbf{Target Acc.}: $77.05_{\pm 0.49}$} & \multicolumn{2}{c}{\textbf{Target Acc.}: $33.57_{\pm 1.66}$}\\
\midrule
2500 & \cellcolor{orange!15}$74.60_{\pm 0.60}$ & \cellcolor{orange!15}$78.96_{\pm 0.58}$ &  \cellcolor{red!15}$32.54_{\pm 1.34}$ & \cellcolor{red!15}$36.96_{\pm 0.99}$\\
6250 & \cellcolor{red!15}$75.86_{\pm 0.23}$ & \cellcolor{red!15}$80.59_{\pm 0.85}$ & \cellcolor{red!15}$35.56_{\pm 1.85}$ & \cellcolor{red!15}$40.49_{\pm 1.43}$\\
12500 & \cellcolor{red!15}$78.29_{\pm 0.51}$ & \cellcolor{red!15}$83.95_{\pm 0.69}$ & \cellcolor{red!15}$35.00_{\pm 3.05}$ & \cellcolor{red!15}$40.11_{\pm 3.20}$\\
25000 & \cellcolor{red!15}$78.00_{\pm 0.30}$ & \cellcolor{red!15}$83.87_{\pm 0.53}$ & \cellcolor{red!15}$34.76_{\pm 1.34}$ & \cellcolor{red!15}$40.47_{\pm 0.87}$\\
\midrule

\multicolumn{5}{c}{\textbf{15\% Poison}}\\
& \multicolumn{2}{c|}{\textbf{Target Acc.}: $75.90_{\pm 0.56}$} & \multicolumn{2}{c}{\textbf{Target Acc.}: $33.51_{\pm 1.95}$}\\
\midrule
2500 & \cellcolor{orange!15}$75.29_{\pm 0.61}$ & \cellcolor{orange!15}$79.16_{\pm 0.22}$ &  \cellcolor{red!15}$33.25_{\pm 1.23}$ & \cellcolor{red!15}$38.20_{\pm 0.87}$\\
 6250 & \cellcolor{red!15}$75.99_{\pm 1.56}$ & \cellcolor{red!15}$81.17_{\pm 1.72}$ & \cellcolor{red!15}$34.23_{\pm 1.71}$ & \cellcolor{red!15}$39.72_{\pm 1.38}$\\
12500 & \cellcolor{red!15}$76.33_{\pm 1.55}$ & \cellcolor{red!15}$81.63_{\pm 0.73}$ & \cellcolor{red!15}$34.89_{\pm 1.11}$ & \cellcolor{red!15}$41.32_{\pm 0.78}$\\
25000 & \cellcolor{red!15}$77.11_{\pm 0.86}$ & \cellcolor{red!15}$82.54_{\pm 1.43}$ & \cellcolor{red!15}$34.99_{\pm 1.63}$ & \cellcolor{red!15}$40.98_{\pm 0.96}$\\
\midrule

\multicolumn{5}{c}{\textbf{20\% Poison}}\\
& \multicolumn{2}{c|}{\textbf{Target Acc.}: $74.70_{\pm 0.40}$} & \multicolumn{2}{c}{\textbf{Target Acc.}: $30.38_{\pm 0.39}$}\\
\midrule
 2500 & \cellcolor{orange!15}$73.03_{\pm 1.32}$ & \cellcolor{orange!15}$76.86_{\pm 1.78}$ & \cellcolor{red!15}$29.78_{\pm 0.57}$ & \cellcolor{red!15}$35.25_{\pm 1.25}$\\
 6250 & \cellcolor{red!15}$75.22_{\pm 0.70}$ & \cellcolor{red!15}$80.20_{\pm 1.74}$ & \cellcolor{red!15}$31.65_{\pm 0.97}$ & \cellcolor{red!15}$37.83_{\pm 1.27}$\\
 12500 & \cellcolor{red!15}$75.41_{\pm 0.70}$ & \cellcolor{red!15}$81.44_{\pm 0.71}$ & \cellcolor{red!15}$31.67_{\pm 1.69}$ & \cellcolor{red!15}$38.34_{\pm 1.34}$\\
25000 & \cellcolor{red!15}$75.65_{\pm 0.95}$ & \cellcolor{red!15}$82.27_{\pm 0.44}$ & \cellcolor{red!15}$31.69_{\pm 1.83}$ & \cellcolor{red!15}$37.51_{\pm 1.17}$\\
\bottomrule

\toprule
\end{tabular}
}
\end{center}
\label{tab:trteeval}
\vspace{-0.75cm} 
\end{table}

%% file: new_experiments.tex
\subsection{Test-time Collusion: \texorpdfstring{\modelext}{ModExt} \texorpdfstring{$\rightarrow$}{Right arrow}\texorpdfstring{\dia}{DistInf}}\label{sec:teeval2}

\change{We validate our guideline’s conjecture on the potential for collusion between \modelext and \dia, marked as ($\uparrow$, $\uparrow$) in Table~\ref{tab:test_interactions}. Here, the common factors are \hyperlink{req3:shadow}{\textbf{\texttt{Te3-Shdw}}} and \hyperlink{req4:architecture}{\textbf{\texttt{Te4-Arch}}} suggesting that \modelext increases the knowledge about $\model$ for better shadow models for \dia. 
We validate by comparing \dia attack accuracy when its shadow models are trained independently on $\daux$ versus extracted from $\model$ using \modelext.}

\noindent\change{\textbf{\underline{Experimental Setup:}}
We assume both \adv{1} and \adv{2} have blackbox access to $\model$. \adv{1} performs \modelext where they steal the functionality of $\model$ by sending queries to $\model$ and using the corresponding outputs to train the surrogate models.

For \modelext, we use KnockoffNets~\cite{orekondy2019knockoff}.
For \dia, we use the blackbox attack from Suri et al.~\cite{suri2023dissecting} where we train an attack classifier to distinguish models trained on datasets with ratios $\alpha_1$ and $\alpha_2$.
For example, the classifier predicts if the proportion of males in the training data is $\alpha_1$ = 0.1 or $\alpha_2$ = 0.9. Following Suri et al.~\cite{suri2023dissecting}, we consider two examples ratios where $\alpha_1$ is close to $\alpha_2$ ($\alpha_1=0.45$ vs. $\alpha_2=0.55$; and $\alpha_1=0.475$ vs. $\alpha_2=0.525$).}

\change{We use two image datasets with sensitive attributes as required for \dia. CELEBA contains 202,599 celebrity face images with 40 binary attributes, and UTKFACE contains 23,705 in-the-wild face images with age, sex, and race attributes. We split each dataset $50/50$ into a train and a test partition, then split each partition $50/50$ (stratified jointly on the label and the sensitive attribute: sex for CELEBA, race for UTKFACE) into target and \adv share.}
\change{We report the attack accuracy of \dia as the mean and standard deviation across five runs. 
% We assume the target model $\model$ is a ResNet34 and consider two settings in which the shadow models either share the same architecture as $\model$ or use a different architecture (VGG11). 
We confirm that fidelity and accuracy of the stolen shadow models are high and omit them for brevity.}
\vspace{-0.5cm}
\begin{table}[!htb]
\caption{\textbf{\modelext $\rightarrow$ \dia:} \emph{Baseline} $\rightarrow$ shadow models trained independently without querying $\model$ (target); \emph{Cross-Arch} $\rightarrow$ VGG11 shadow model extracted from $\model$; \emph{Same-Arch} $\rightarrow$ ResNet34 shadow model extracted from $\model$. \colorbox{green!15}{Green} $\rightarrow$ higher than baseline, \colorbox{orange!15}{orange} $\rightarrow$ similar to baseline (within std.\ dev.), \colorbox{red!15}{red} $\rightarrow$ lower than baseline.}
\label{tab:modextDIA}
\vspace{0.25cm}
\resizebox{\columnwidth}{!}{
\begin{tabular}{ @{\hspace{6pt}} l @{\hspace{4pt}} | @{\hspace{4pt}} c @{\hspace{4pt}}  @{\hspace{4pt}} c @{\hspace{4pt}} | @{\hspace{4pt}} c @{\hspace{4pt}}  @{\hspace{4pt}} c @{\hspace{4pt}}  }
\bottomrule

\toprule
\multirow{3}{*}{\textbf{Setting}} & \multicolumn{2}{c}{\textbf{CELEBA}} & \multicolumn{2}{c}{\textbf{UTKFACE}} \\
 & $\alpha_1=0.45$ & $\alpha_1=0.475$ & $\alpha_1=0.45$ & $\alpha_1=0.475$ \\
 & $\alpha_2=0.55$ & $\alpha_2=0.525$ & $\alpha_2=0.55$ & $\alpha_2=0.525$ \\
\midrule
\textbf{Baseline.} & $70.00_{\pm 14.58}$ & $55.00_{\pm 18.11}$ & $55.00_{\pm 3.06}$ & $45.50_{\pm 10.81}$ \\
\textbf{Cross-Arch} & \cellcolor{green!15}$98.00_{\pm 3.26}$ & \cellcolor{green!15}$96.50_{\pm 4.18}$ & \cellcolor{green!15}$91.50_{\pm 4.54}$ & \cellcolor{green!15}$88.00_{\pm 7.79}$ \\
\textbf{Same-Arch} & \cellcolor{green!15}$99.00_{\pm 1.37}$ & \cellcolor{green!15}$98.50_{\pm 3.35}$ & \cellcolor{green!15}$93.00_{\pm 7.37}$ & \cellcolor{green!15}$92.50_{\pm 4.68}$ \\
\bottomrule

\toprule
\end{tabular}
}
\vspace{-0.25cm}
\end{table}

\noindent\change{\textbf{\underline{Results:}} We report three settings in Table~\ref{tab:modextDIA}: \emph{baseline} where the shadow models for \dia are trained independently without querying $\model$; \emph{Cross-Arch} where the shadow models have a different architecture than $\model$ (i.e., VGG11) and obtained using \modelext; and \emph{Same-Arch} where the shadow model has the same architecture as $\model$, obtained using \modelext.}

\change{As expected, we find that the attack accuracy of \dia is significantly higher for cross-arch and same-arch settings (where shadow models are obtained using \modelext) compared to the case where the shadow models are trained independently on $\daux$ (in \colorbox{green!15}{green}). This confirms our conjecture that \modelext improves \dia.}

\subsection{Test-time Collusion: \texorpdfstring{\datarecon}{DataRecon} \texorpdfstring{$\rightarrow$}{right arrow} \{\texorpdfstring{\mia}{MemInf}, \texorpdfstring{\aia}{AttInf}, \texorpdfstring{\dia}{DistInf}\}}\label{sec:teeval3}

\change{We validate our conjecture on collusion between \datarecon and three privacy attacks: \mia, \aia, and \dia, marked as ($\uparrow$, $\uparrow$) in Table~\ref{tab:test_interactions}. The common factor is \hyperlink{req1:aux}{\textbf{\texttt{Te1-Aux}}} suggesting that \datarecon improves the quality of $\daux$ to be similar to $\dtrain$. This helps train better shadow models for \mia, \aia, and \dia. We validate the conjecture by comparing the attack accuracy of \mia, \aia, and \dia in two settings: (i) when $\daux$ has no overlap with $\dtrain$, and (ii) when $\daux$ includes records extracted using \datarecon, and used to train the shadow models. We evaluate using the mean and standard deviation over five runs unless stated otherwise.}

\noindent\change{\textbf{\underline{Experimental Setup:}} To validate the conjecture, we generate reconstructed data by executing \datarecon on a fraction of $\dtrain$. We then replace the fraction of $\daux$ with these reconstructed $\dtrain$ records (aka replacement ratio, we use \{25\%, 50\%, 75\%\}). For \datarecon, we use generative gradient inversion~\cite{gifd}. We assume \adv{1} has whitebox access to $\model$ (e.g., in a federated learning setting where \adv can access the model gradients), and uses this to reconstruct data records in $\dtrain$. These reconstructed data records are then shared with \adv{2} to train shadow models for \mia, \aia, and \dia.
}

\change{For \mia, we use the likelihood ratio attack~\cite{carlini2022membership} on CIFAR10/100 (\S\ref{sec:trteeval}). Each dataset’s 50k training images are split equally into a target pool (members) and \adv's non-member pool ($\daux$). $\model$ is trained on a random 12.5k subset of the target pool, while \adv trains 64 shadow models (50 epochs each) on random halves of a 15k-record $\daux$. $\dtest$ is the standard 10k records from the datasets. We evaluate on a fixed disjoint set of 2k records (1k members, 1k non-members). Overall, we have $|\dtrain|=12{,}500, |\dtest|=10{,}000, |\daux|=25{,}000$, and a membership evaluation set of size $2{,}000$. All models are ResNet34.}

\change{For \aia, we use the attack from Aalmoes et al.~\cite{aalmoes2025alignment} on CELEBA and UTKFACE (\S\ref{sec:teeval3}). For CELEBA, the target task is smile prediction and the sensitive attribute is sex; for UTKFACE, the target task is predicting sex and the sensitive attribute is race. Stratified on the label and the sensitive attribute, the target trains on $\dtrain$ ($1{,}000$ members, of which \adv reconstructs $500$ and the other $500$ are held-out), \adv holds $\daux$ ($1{,}500$ non-members), and evaluation queries the $500$ held-out members and $500$ held-out non-members. Overall, we have $|\dtrain|=1{,}000, |\daux|=1{,}500$, and an evaluation set of size $1{,}000$. Both $\model$ and attack model use ResNet18.}

\change{Finally, for \dia, we use the blackbox attack of Suri et al.~\cite{suri2023dissecting} on CELEBA and UTKFACE (\S\ref{sec:teeval2}), with sex as the sensitive attribute. $\model$ is trained at ratio $\alpha_1$, and the attack asks whether $\model$'s ratio is $\alpha_1$ or an alternative ratio $\alpha_2$. We train 128 shadow models (64 with subgroup ratio $\alpha_1$, 64 with $\alpha_2$), each on 200 records for 12 epochs. Through each shadow model, we pass a fixed set of $500$ query images (half male, half female), and record its loss on each which are used by a logistic-regression attack classifier to predict whether $\model$ was trained with $\alpha_1$ or $\alpha_2$. We report average accuracy over three runs.
$\model$ and shadow models are ResNet18.}

\noindent\change{\textbf{\underline{Results (\datarecon $\rightarrow$ \mia):}} We report the TPR@$0.01$ FPR for different percentages of replaced records. For the first dataset, the baseline TPR of $1.50$ increases to $2.50$ at $25\%$ replacement and further to $3.50$ at $50\%$, before dropping to $2.10$ at $75\%$ replacement. Similarly, for the second dataset, the baseline TPR of $2.50$ improves to $2.70$ at $25\%$ replacement and $2.80$ at $50\%$, but decreases to $2.60$ at $75\%$ replacement. Overall, the TPR@$0.01$ FPR improves with moderate levels of replacement, while a higher replacement ratio of $75\%$ leads to a decline in TPR for both datasets. We conjecture that the benefit of adding member-like records to the shadow training set saturates after a few hundred records, while each inexact reconstruction acts as a slightly mislabeled member whose error accumulates as more are replaced. 
Thus, for moderate ratios, we find that \datarecon can improve \mia, confirming our conjecture for the collusion potential.}

\noindent\change{\textbf{\underline{Results (\datarecon $\rightarrow$ \aia):}} We report the AUC score under ROC curve. 
For CELEBA, the baseline AUC at $0\%$ replacement is $69.25\%$ (average value), which increases to $71.45\%$ at $25\%$ replacement, $73.07\%$ at $50\%$, and reaches $73.95\%$ at $75\%$ replacement. Similarly, for UTKFACE, the baseline AUC of $62.94\%$ improves consistently with higher replacement ratios, increasing to $64.85\%$ at $25\%$, $66.03\%$ at $50\%$, and $67.42\%$ at $75\%$ replacement. We find that the improvement for $50\%$ and $75\%$ replacement gives significant performance improvement when compared to the baseline. Overall, for both datasets, increasing the percentage of replaced records leads to an improvement in AUC score.
This confirms our conjecture that there is potential for collusion where \datarecon can help improve \aia.}

% We see that increasing the \% of replaced data records to 50\% and 75\% results in higher AUC score for both datasets, while balanced accuracy improves in the case of 

% \begin{table}[!htb]
% \caption{\textbf{\datarecon} $\rightarrow$ \textbf{\aia}: We report balanced accuracy and AUC score and use \colorbox{green!15}{green} $\rightarrow$ higher than baseline, \colorbox{orange!15}{orange} $\rightarrow$ similar to baseline, \colorbox{red!15}{red} $\rightarrow$ lower than baseline.}
% \begin{center}
% \resizebox{0.85\columnwidth}{!}{
% \begin{tabular}{ l|c|c|c|c }
% \bottomrule

% \toprule
% \multirow{2}{*}{\textbf{Replaced \%}} & \multicolumn{2}{c|}{\textbf{CELEBA}} & \multicolumn{2}{c}{\textbf{UTKFACE}} \\
% & \textbf{Bal.\ Acc} & \textbf{AUC} & \textbf{Bal.\ Acc} & \textbf{AUC} \\
% \bottomrule

% \toprule
% \textbf{Baseline (0\%)} & $61.22_{\pm4.47}$ & $69.25_{\pm4.18}$ & $57.56_{\pm3.74}$ & $62.94_{\pm2.81}$ \\

% \textbf{25\%} & \cellcolor{orange!15}$61.39_{\pm3.35}$ & \cellcolor{orange!15}$71.45_{\pm3.13}$ & \cellcolor{orange!15}$59.84_{\pm2.39}$ & \cellcolor{orange!15}$64.85_{\pm2.79}$ \\

% \textbf{50\%} & \cellcolor{orange!15}$62.42_{\pm3.78}$ & \cellcolor{green!15}$73.07_{\pm3.38}$ & \cellcolor{green!15}$61.11_{\pm2.19}$ & \cellcolor{green!15}$66.03_{\pm1.59}$ \\

% \textbf{75\%} & \cellcolor{orange!15}$64.47_{\pm3.60}$ & \cellcolor{green!15}$73.95_{\pm3.47}$ & \cellcolor{green!15}$60.77_{\pm1.44}$ & \cellcolor{green!15}$67.42_{\pm1.62}$ \\
% \bottomrule

% \toprule
% \end{tabular}
% }
% \end{center}
% \label{tab:teeval_aia}
% \end{table}

\noindent\change{\textbf{\underline{Results (\datarecon $\rightarrow$ \dia):}} We report the attack accuracy results for \dia under two ratio configurations: $\alpha_1{=}0.5$ vs.\ $\alpha_2{=}0.1$, and $\alpha_1{=}0.5$ vs.\ $\alpha_2{=}0.9$ (following results in Suri and Evans~\cite{suri2022formalizing}). For UTKFACE with $\alpha_2{=}0.1$, the baseline attack accuracy at $0\%$ replacement is $62.66\%$, which increases to $74.06\%$ at $25\%$, $\sim76\%$ at $50\%$ and $75\%$ replacement. For $\alpha_2{=}0.9$, the attack accuracy improves from a baseline of $60.31\%$ to $72.50\%$ at $25\%$, $72.81\%$ at $50\%$, and further to $75.31\%$ at $75\%$ replacement.}

\change{For CELEBA with $\alpha_2{=}0.1$, the baseline attack accuracy of $60.47\%$ increases consistently to $64.53\%$ at $25\%$ replacement, $68.91$ at $50\%$, and $71.09\%$ at $75\%$. The strongest trend is observed for CELEBA with $\alpha_2{=}0.9$, where the baseline accuracy of $63.44\%$ increases substantially to $80.78\%$ at $25\%$ replacement, $82.81\%$ at $50\%$, and reaches $87.66\%$ at $75\%$ replacement. Overall, across both datasets and ratio configurations, increasing the percentage of reconstructed records in $\daux$ consistently improves \dia, supporting the conjecture that \datarecon can increase collusion potential.}

%% file: 6capability.tex
\section{Accounting for Adversary's Characteristics}\label{sec:characteristics}

So far, we did not explicitly take \adv's characteristics into account but implicitly assumed that each \adv possesses the minimal set of characteristics needed to execute the attack. 
We now extend our framework to consider \adv's characteristics by discussing the impact of changing them on the collusion potential.
\change{We present Tables~\ref{tab:trends_attackone} and~\ref{tab:trends_attacktwo} showing the impact of collusion potential on varying \adv's characteristics. 
The notations for \adv's characteristics are based on our proposed streamlined threat model in Appendix~\ref{sec:streamlined}. 
The notations are summarized in Table~\ref{tab:summary}. 
We consider the following two cases depending on which \adv is changed:}
\vspace{-0.25cm}
\input{tables/tab_adv_desc}

\input{tables/tab_trends}

\begin{itemize}[leftmargin=*]
    \item \cosemtic{\underline{\textbf{\adv{1} $\rightarrow$ Change; \adv{2} $\rightarrow$ Same}}: Table~\ref{tab:trends_attackone} shows the minimum characteristics for each risk that we implicitly assumed so far (indicated as \minimum).
    Increasing \adv{1}'s characteristics will increase the effectiveness of \attackone and the correlations with underlying factors (Tables~\ref{tab:train_test_relation} and~\ref{tab:test_time_relation}). 
    This, in turn, increases the potential for collusion and facilitates \attacktwo. 
    From Table~\ref{tab:trends_attackone}, we can determine whether the potential for collusion increases based on increasing \adv{1}'s characteristics beyond the minimum requirements (indicated as \trend) when going from left (weak) to right (strong).}

    \item \cosemtic{\underline{\textbf{\adv{1} $\rightarrow$ Same; \adv{2} $\rightarrow$ Change}}: 
    Recall that the guideline identifies a potential for collusion when the \emph{factors} required for \attacktwo are satisfied by the \emph{outcomes} of \attackone. Strengthening \adv{2}'s characteristics can increase the effectiveness of \attacktwo, though not necessarily by increasing collusion potential. In some cases, stronger characteristics may directly satisfy prerequisites required for \attacktwo, independent of \attackone's outcomes, thereby reducing or eliminating the need for collusion. Therefore, collusion should be evaluated only over the factors for \attacktwo that remain unmet after accounting for \adv{2}'s characteristics. Table~\ref{tab:trends_attacktwo} summarizes such cases, where \partialColl indicates reduced collusion potential under a specific characteristic, and \xmark indicates that collusion with \attackone becomes unnecessary. We illustrate how strengthening \adv{2}'s characteristics removes or reduces the need for collusion, using Table~\ref{tab:trends_attacktwo}.}
\begin{itemize}[leftmargin=*]
    \item \change{\textbf{Model knowledge:} Granting \adv{2} whitebox access to $\model$ (\whitebox) satisfies both the functionality and architecture requirements (\hyperlink{req3:shadow}{\textbf{\texttt{Te3-Shdw}}}, \hyperlink{req4:architecture}{\textbf{\texttt{Te4-Arch}}}), which \modelext\ would otherwise provide. As a result, attacks such as \mia, \aia, \dia, and \datarecon can be carried out effectively \emph{without} colluding with \adv{1} for \modelext\ (\xmark\ in Table~\ref{tab:trends_attacktwo}). Granting only graybox access (\graybox) satisfies the architecture requirement (\hyperlink{req4:architecture}{\textbf{\texttt{Te4-Arch}}}) but not functionality; this \emph{reduces} but does not remove the need for collusion with \adv{1} for \modelext\ (\partialColl).}
    
    \item \change{\textbf{Data knowledge ($\daux$):} \mia, \evasion, and \datarecon produce $\daux$ as an outcome (\hyperlink{req1:aux}{\textbf{\texttt{Te1-Aux}}}), which can strengthen other attacks. However, \hyperlink{req1:aux}{\textbf{\texttt{Te1-Aux}}} may be satisfied directly by increasing \adv{2}'s knowledge about $\dtrain$: full overlap (\fullOverlap) removes the need to collude with \adv{1} for \mia, \evasion, and \datarecon\ (\xmark), while partial overlap (\partialOverlap) reduces it (\partialColl).}
    
    \item \change{\textbf{Data knowledge (distribution):} \aia\ and \dia\ provide $\dtrain$'s distribution as an outcome (\hyperlink{req2:dist}{\textbf{\texttt{Te2-Dist}}}), which strengthens attacks that require \hyperlink{req2:dist}{\textbf{\texttt{Te2-Dist}}}, such as \aia, \evasion, \mia, \datarecon, and \modelext. \hyperlink{req2:dist}{\textbf{\texttt{Te2-Dist}}} can instead be satisfied by increasing \adv{2}'s knowledge about $\dtrain$: \fullOverlap\ removes the need for collusion while \partialOverlap\ reduces it.}
\end{itemize}
\end{itemize}

\setlength{\tabcolsep}{4pt}
\begin{table}[ht]
\centering
\caption{\change{\textbf{\ul{Impact of \adv's Characteristics on Collusion Potential (\attacktwo):}}
Rows correspond to \attacktwo; we vary \adv{2}'s characteristics while \adv{1} is fixed.
Strengthening \adv{2} raises effectiveness directly (not via collusion, hence omitted); we instead mark its effect on collusion potential with the listed \attackone(s):
\partialColl $\rightarrow$ collusion potential decreases with characteristic;
\xmark $\rightarrow$ collusion not required as characteristic meets the prerequisite factor. Columns under \knowledgeModelSet are left empty for \modelext, as they are the outcomes of \modelext.}}
\label{tab:trends_attacktwo}
\vspace{0.25cm}
\resizebox{\columnwidth}{!}{
\begin{tabular}{l||c|cc|cc|cc}
\bottomrule

\toprule
& \multicolumn{3}{c|}{\textbf{\knowledgeDataSet}} & \multicolumn{3}{c}{\textbf{\knowledgeModelSet}} \\
\textbf{\attacktwo} & 
\textbf{\attackone} & 
\rotatebox{90}{\textbf{\partialOverlap}} & 
\rotatebox{90}{\textbf{\fullOverlap}} & 
\textbf{\attackone} & 
\rotatebox{90}{\textbf{\graybox}} & 
\rotatebox{90}{\textbf{\whitebox}} \\
\bottomrule

\toprule
\textbf{\evasion} & \mia, \datarecon, \aia, \dia & \partialColl & \xmark & \modelext & \partialColl & \xmark \\
\midrule
\textbf{\modelext} & \mia, \evasion, \datarecon, \aia, \dia & \partialColl & \xmark & --- & --- & ---\\
\midrule
\textbf{\mia} & \evasion, \datarecon, \aia, \dia & \partialColl & \xmark & \modelext & \partialColl & \xmark \\
\midrule
\textbf{\aia} & \mia, \evasion, \datarecon, \dia & \partialColl & \xmark & \modelext & \partialColl & \xmark \\
\midrule
\textbf{\dia} & \mia, \evasion, \datarecon, \aia & \partialColl & \xmark & \modelext & \partialColl & \xmark \\
\midrule
\textbf{\datarecon} & \mia, \evasion, \aia, \dia & \partialColl & \xmark & \modelext & \partialColl & \xmark \\
\bottomrule

\toprule
\end{tabular}
}
\vspace{-0.5cm}
\end{table}

\noindent Based on the above discussion, we need to refine the guideline to account for the cases where strengthening \adv{2}'s characteristics satisfies \attacktwo’s prerequisites without requiring collusion. Specifically, after identifying common factors between \attackone and \attacktwo from Tables~\ref{tab:train_test_relation} and~\ref{tab:test_time_relation}, we use Tables~\ref{tab:trends_attackone} and~\ref{tab:trends_attacktwo} to determine whether \adv{2}'s characteristics already meet the prerequisites for \attacktwo. 
Hence, when assessing collusion potential, we refine our guideline (\textbf{\ref{guideline:G1}}) to consider only the common \emph{unmet} factors between risks \emph{after} accounting for the adversary’s characteristics.

\begin{takeaway}
    \textbf{Takeaway:} We revise our guideline to account for the setting where \adv{2} has characteristics which meet the pre-requisite, removing the need for collusion.
\end{takeaway}

%% file: tables/tab_adv_desc.tex
\setlength{\tabcolsep}{4pt}
\begin{table}[!htb]
\caption{\textbf{\adv's Characteristics and their descriptions}. Detailed descriptions provided in Appendix~\ref{sec:streamlined}.}
% \begin{center}
\vspace{0.25cm}
\resizebox{0.95\columnwidth}{!}{
\begin{tabular}{ l|l|p{4.15cm} } 
 \bottomrule

 \toprule
 \textbf{Category} & \textbf{Notation} & \textbf{Description} \\ 
 \bottomrule

 \toprule
 \multirow{8}{*}{\textbf{Role} (\advSet)} & \dataOwner & Data Owner \\
 & \dataProvider & Data Provider \\
 & \codeProvider & Code Provider \\
 & \modelProvider & Model Provider \\
 & \modelTrainer & Model Trainer \\
 & \modelOwner & Model Owner \\
 & \serviceProvider & Service Provider \\
 & \client & Client \\
 \midrule
 \multirow{9}{*}{\textbf{Objective} (\objectiveSet)} & \evasion & Evasion \\
 & \poison & Poisoning \\
& \backdoor & Backdoor\\
& \mia & Membership Inference\\
& \aia & Attribute Inference\\
& \dia & Distribution Inference\\
& \datarecon & Data Reconstruction\\
& \modelext & Unauth. Model Ownership\\
& \disc & Discriminatory Behavior\\
 \midrule
 \multirow{2}{*}{\textbf{Capability} (\capabilitySet)} & \passiveAdv & Honest-but-Curious \adv \\
& \activeAdv & Malicious \adv \\
 \midrule 
 \multirow{2}{*}{\textbf{Optimization} (\optimizationSet)} & \adaptive & Adaptive\\
 & \nonAdaptive & Non-Adaptive\\
\midrule
 \multirow{7}{*}{\textbf{Model Knowledge} (\knowledgeModelSet)} 
 & \noAccess & No Access \\
 & \blackbox & Blackbox\\
& \graybox & Graybox\\
& \whitebox & Whitebox\\\cline{2-3}
& \hardLabel & Hard labels\\
& \topK & Top-K Predictions\\
& \fullPrediction & Full prediction vector\\
\midrule
 \multirow{4}{*}{\textbf{Interaction} (\interactionSet)} & \noAccess & No Access \\
 & \oneShot & One-Shot (Single Query)\\
& \kShot &  K-Shot ($k$ Queries) \\
& \unlimited & Unlimited Queries\\
\midrule
 \multirow{6}{*}{\textbf{Data Knowledge} (\knowledgeDataSet)} & \noAccess & No Access \\
& \noOverlap & No-overlap with $\dtrain$\\
& \partialOverlap & Partial overlap\\
& \fullOverlap & Full overlap\\\cline{2-3}
& \noBlind & Non-Blind \\
& \blind & Blind \\
% \midrule
%  \multirow{3}{*}{\textbf{Train-test Factors}} & \hyperlink{tr1:tail}{\textbf{\texttt{Tr1-Tail}}} & Tail length of $\dtrain$'s distribution \\  
% & \hyperlink{tr2:dataset}{\textbf{\texttt{Tr2-Data}}} & Distinguishability across datasets\\ 
% & \hyperlink{tr3:subgroups}{\textbf{\texttt{Tr3-Subgp}}} & Distinguishability across subgroups \\
% \midrule
%  \multirow{4}{*}{\textbf{Test-time Factors}} & \hyperlink{req1:aux}{\textbf{\texttt{Te1-Aux}}} & Auxiliary Dataset \\ 
% & \hyperlink{req2:dist}{\textbf{\texttt{Te2-Dist}}} & Data Distribution \\ 
% & \hyperlink{req3:shadow}{\textbf{\texttt{Te3-Shdw}}} & Model Functionality \\ 
% & \hyperlink{req4:architecture}{\textbf{\texttt{Te4-Arch}}} & Model Architecture \\ 
% \midrule
% \multirow{3}{*}{\textbf{General Notations}} & $\model$ & Target Model\\
%  & $\surrogatemodel$ & Surrogate Model\\
%  & $\dtrain$, $\dtest$, $\daux$ & Train, test, auxiliary data\\
 \bottomrule

 \toprule
\end{tabular}
}
% \end{center}
\label{tab:summary}
\vspace{-0.25cm}
\end{table}

%% file: tables/tab_trends.tex
\setlength{\tabcolsep}{4pt}
\begin{table*}[!htb]
\caption{\change{\textbf{\underline{Impact of \adv's Characteristics on Collusion Potential (\adv{1} $\rightarrow$ Change; \adv{2} $\rightarrow$ Same):}}
Rows correspond to \attackone; we vary \adv{1}'s characteristics while \adv{2} is fixed; values range from weak (left) to strong (right).
\minimum $\rightarrow$ minimum requirement for \adv{1} to execute \attackone;
\trend $\rightarrow$ collusion potential with \attacktwo increases by strengthening the characteristic from \minimum; empty cell $\rightarrow$ insufficient, unrealistic, or unnecessary. Refer to Appendix~\ref{sec:streamlined} for description of notations. The table markings are based on prior work covering each of the attacks~\cite{SoKMLPrivSec,suriSoK,FenauxSoK,poisonSurvey2,modelextSurvey,meminfSurvey}.}}
\begin{center}
\resizebox{0.8\textwidth}{!}{
\begin{tabular}{l|c|cc|cc|cccc|ccc|cccc|cccc|cc}
\bottomrule

\toprule
\textbf{\attackone} & \textbf{\advSet} & \multicolumn{2}{c|}{\textbf{\capabilitySet}} & \multicolumn{2}{c|}{\textbf{\optimizationSet}} & \multicolumn{7}{c|}{\textbf{\knowledgeModelSet}} & \multicolumn{4}{c|}{\textbf{\interactionSet}} & \multicolumn{6}{c}{\textbf{\knowledgeDataSet}} \\
\midrule
& &  \rotatebox{90}{\textbf{\passiveAdv}} & \rotatebox{90}{\textbf{\activeAdv}} & \rotatebox{90}{\textbf{\nonAdaptive}} & \rotatebox{90}{\textbf{\adaptive}}  & \rotatebox{90}{\textbf{\noAccess}} & 
 \rotatebox{90}{\textbf{\blackbox}} & \rotatebox{90}{\textbf{\graybox}} & \rotatebox{90}{\textbf{\whitebox}}  &  \rotatebox{90}{\textbf{\hardLabel}}  & \rotatebox{90}{\textbf{\topK}} & \rotatebox{90}{\textbf{\fullPrediction}} & 
 \rotatebox{90}{\textbf{\noAccess}}  & \rotatebox{90}{\textbf{\oneShot}} & \rotatebox{90}{\textbf{\kShot}} & 
 \rotatebox{90}{\textbf{\unlimited}} & 
 \rotatebox{90}{\textbf{\noAccess}} & 
 \rotatebox{90}{\textbf{\noOverlap}} & \rotatebox{90}{\textbf{\partialOverlap}} & \rotatebox{90}{\textbf{\fullOverlap}} & \rotatebox{90}{\textbf{\blind}} & \rotatebox{90}{\textbf{\noBlind}} \\
\bottomrule

\toprule
% \multicolumn{23}{c}{\textbf{\underline{Train-time Risk}}} \\

\multirow{4}{*}{\textbf{\poison}} & \dataOwner/\dataProvider & & \minimum & \minimum  & & \minimum & & & & & & & \minimum & & & & & & \minimum & \trend & \minimum & \trend \\
& \modelTrainer & & \minimum & \minimum & \trend &  &  &  & \minimum &  &  & \minimum &  &  &  & \minimum &  & & & \minimum &  & \minimum \\
& \modelProvider & & \minimum & \minimum & &  &  & \minimum & &  &  &  &  &  &  & & \minimum & & & & &  \\
& \codeProvider & & \minimum & \minimum & & \minimum & & & &  &  &  & \minimum & & & & \minimum & & & & &  \\
\midrule

% \multicolumn{23}{c}{\textbf{\underline{Test-time Risks}}} \\
\textbf{\evasion} & \client &  & \minimum & \minimum & \trend &   & \minimum  & \trend & \trend & \minimum & \trend & \trend &  & \minimum & \trend & \trend & & \minimum & \trend & \trend & \minimum & \trend \\
\midrule
\textbf{\modelext} & \client & \minimum & & \minimum & \trend &  & \minimum &  & & \minimum & \trend & \trend &  &  & \minimum & & & \minimum & \trend & \trend & \minimum & \trend\\
\midrule
\textbf{\mia} & \client & \minimum & & \minimum & \trend   &  & \minimum & \trend & \trend & \minimum & \trend & \trend &  & \minimum & \trend & \trend & & \minimum & \trend & \trend & \minimum & \trend\\
\midrule
\textbf{\aia} & \client & \minimum & & \minimum & \trend   &  & \minimum & \trend & \trend & \minimum & \trend & \trend &  & \minimum & \trend & \trend & & \minimum & \trend & \trend & \minimum & \trend\\
\midrule
\textbf{\dia} & \client & \minimum & & \minimum & \trend   &  & \minimum & \trend & \trend & \minimum & \trend & \trend &  & \minimum & \trend & \trend & & \minimum & \trend & \trend &  & \minimum\\
\midrule
\textbf{\datarecon} & \client & \minimum & & \minimum & \trend   &  & \minimum & \trend & \trend & \minimum & \trend & \trend &  & \minimum & \trend & \trend & & \minimum & \trend & \trend & \minimum & \trend\\
\bottomrule

\toprule
\end{tabular}
}
\end{center}
\label{tab:trends_attackone}
\vspace{-0.5cm}
\end{table*}

%% file: 8discussions.tex
\section{Discussion and Summary}\label{sec:discussion}

% We show that our framework is \emph{comprehensive} and \emph{extensible} to new factors and risks (\Snospace\ref{sec:meetreqs}).
% We further show how our guideline extends beyond two colluding adversaries along with some exceptions to our guideline (\Snospace\ref{sec:extending}).
% Finally, we discuss how multiple colluding adversaries can be represented as a single entity (\Snospace\ref{sec:multiple_adv}).

% \subsection{Framework and Requirements}\label{sec:meetreqs}

\noindent\textbf{\underline{Framework and Requirements:}}
We evaluate our framework against the two stated requirements: comprehensiveness and extensibility.

\noindent\textbf{\ref{comprehensive} Comprehensive:}
Our framework is comprehensive with respect to both the risks considered for collusion and the factors used to relate them.

\begin{itemize}[leftmargin=*]
    \item \emph{Comprehensiveness of risks:} 
    Our framework covers security, privacy, and fairness risks that are sufficiently studied in prior work to support systematic reasoning about collusion across the ML pipeline. These include both train-time and test-time risks for which interactions can be characterized through shared underlying factors. 
    
    Risks that lack sufficient prior analysis to enable such characterization are outside the scope of this work. For example:
    \begin{itemize}[leftmargin=*]
        \item \emph{Side-channel attacks~\cite{batina2019csinn, yan2020cache, hua2018reverse}} operate under a blackbox $\model$ (\blackbox) where a \client or \serviceProvider exploits physical or system-level leakage to extract model architecture, and partial model parameters.
        
        \item \emph{Fault injection~\cite{yao2020deephammer, rakin2021deepdup, lao2022deepstrike}} occurs when a \client or \serviceProvider exploits hardware vulnerabilities to induce targeted faults during execution or weight storage, enabling accuracy degradation or targeted misclassification.
        
        \item \emph{Adversarial initialization~\cite{grosse2019init, grosse2020advinit}} is a train-time attack where a \modelProvider or \modelTrainer initialize the model such that standard training on benign data converges to \adv-desired behavior (e.g., backdoors, increased training time, or biased outputs).
        
        \item \emph{Data ordering attacks~\cite{shumailov2021dataOrdering, anonymous2026toga}} is a specific form of \poison in which a \dataOwner or \dataProvider poisons the model by reordering data during training.
        
        \item \emph{Sponge examples~\cite{shumailov2021sponge, cina2025sponge, wang2023sponge}} violates availability where \client sends carefully crafted inputs that increase energy consumption and latency.
    \end{itemize}
    As research on these risks matures, they can be incorporated into our framework by reasoning about \adv's characteristics and any additional factors required for the risk.

    \item \emph{Comprehensiveness of factors:} 
    Our framework identifies a set of train-test and test-time factors that underlie collusion among risks. These factors are drawn from prior work and are common across the risks considered in this paper (\Snospace\ref{sec:background}). 
    We show how \emph{each risk} relates to \emph{all} identified factors, thus enabling systematic reasoning about collusion (Tables~\ref{tab:train_test_relation} and~\ref{tab:test_time_relation}). While we do not claim that the identified factors are complete, our framework allows adding more factors.
\end{itemize}

\noindent Overall, our framework is comprehensive within its defined scope of risks in \Snospace\ref{sec:background} (meets \ref{comprehensive}) and can be extended to incorporate new risks and factors.

\noindent\textbf{\ref{extensible} Extensible:}
Our framework is designed to support additional factors, risks, and adversary characteristics.
To add new factors, we can specify how each factor relates to existing risks, as illustrated in Tables~\ref{tab:train_test_relation} and~\ref{tab:test_time_relation}.
For new risks, we characterize how the risk relates to the existing factors (e.g., Tables~\ref{tab:train_test_relation} and~\ref{tab:test_time_relation} and \Snospace\ref{sec:revisitRisks}). 
Once these relations are defined, the guideline can be used to study the potential collusion.
% Below we discuss how the guideline itself can be extended to reason about collusion beyond two adversaries (\Snospace\ref{sec:extending}). 
Overall, our framework satisfies the extensibility requirement (\ref{extensible}).

\begin{takeaway}
    \textbf{Takeaway:} Our proposed framework is comprehensive (\ref{comprehensive}) and extensible (\ref{extensible}).
\end{takeaway}

\noindent\textbf{\underline{Extending Beyond Two Adversaries:}}
% \subsection{Extending Beyond Two Adversaries}\label{sec:extending}
Our framework extends naturally to collusions with more than two adversaries. We first discuss two cases where \adv{1} colludes with multiple adversaries in \emph{parallel} (we use ``$\Rightarrow$'' for ``colludes with''):
\begin{itemize}[leftmargin=*]
    \item \textbf{Train \adv{1} $\Rightarrow$ Multiple Test \adv{s}}\\ 
    \adv{1} can optimize train-time risks to increase susceptibility to multiple test-time risks, enabling collusion with several test-time adversaries.

    \item\textbf{Test \adv {1} $\Rightarrow$ Multiple Test \adv{s}}\\ 
    \adv{1} executes a test-time attack and uses the knowledge learned as a prerequisite to increase the effectiveness of other test-time attacks. This will allow \adv{1} to collude with multiple test-time adversaries.
\end{itemize}

For the above two cases, we can apply our guideline (from \Snospace\ref{sec:guideline}) between \adv{1} and each of the other test-time adversaries in parallel.
The potential for overall collusion will be determined independently by the potential for each pairwise collusion.
We now present a case for \emph{sequential collusion}:
\begin{itemize}[leftmargin=*]        
    \item\textbf{Test or Train \adv{1} $\Rightarrow$ Test \adv{2} $\Rightarrow$ Test \adv{3} $\Rightarrow$ $\cdots$} \\
\adv{1} colludes with \adv{2} to increase effectiveness of \attacktwo executed by \adv{2}, whose success provides knowledge enabling downstream test-time collusion.
\end{itemize}
Here, we can apply our guideline sequentially to each pairwise collusion.
The potential for overall collusion will be determined by the potential for pairwise collusions.
More complex combinations of the parallel and sequential cases may exist, but our guideline can be readily extended to predict potential collusion.

\noindent\change{\textbf{\underline{Adaptive Strategies:}} One possible setting involves multi-round collusion, where \adv{1} iteratively adapts their attack strategy based on the success of \adv{2}'s attack in the previous round. In this scenario, the effectiveness and collusion potential are expected to increase across rounds. Specifically, improvements in \attackone by \adv{1} based on feedback from previous round, will further improve the effectiveness of \attacktwo executed by \adv{2}. This iterative amplification is captured by our guideline through an increase in correlation between \attackone and the factors across successive rounds, consequently explaining the increase in effectiveness of \attacktwo.}

\noindent\textbf{\underline{Exceptions to Guideline:}} 
We do not claim our framework is perfect; rather, it is a first step toward systematically understanding and predicting the potential for collusion among adversaries.
Our guideline cannot assess collusion when there are no common factors. While we did not encounter any such cases, further work would be needed to either uncover more relations between risks and factors to apply the guideline, or identify additional missing factors that can enable the collusion.
This can then allow us to apply our guideline to evaluate collusion among adversaries.

\noindent\change{\textbf{\underline{Application of Framework and Guideline:}} Various stakeholders in the ML pipeline can be impacted by the increased effectiveness of colluding adversaries. For instance, \emph{data owners}, \emph{model owners}, and \emph{service providers} become more susceptible to security, privacy, and fairness risks under collusion. Our framework helps identify possible sources of collusion, the enabling factors, and the potential impact of such collusion.
\emph{ML security researchers and practitioners} can use our framework to study previously unexplored collusion settings, develop stronger attacks for auditing, design defenses, and analyze enabling factors. Practitioners can further account for these factors to reduce collusion potential (see below), and use our guideline to efficiently assess collusion risks.}

\noindent\change{\underline{\textbf{Potential Mitigations:}} By explicitly identifying the enabling factors through our framework and guideline, our work can help researchers and practitioners design defenses that disrupt collusion. For train-test collusion, defenses such as detecting and filtering poisoned samples in $\dtrain$~\cite{soremekun2023towards}, modifying training objectives to reduce the influence of poisons~\cite{zhao2024adversarially}, can reduce the impact of \poison that enable subsequent attacks. Additionally, specific factors can be explicitly targeted to reduce correlations (e.g., reduce subgroup distinguishability using fairness constraints~\cite{aalmoes2025alignment}). For test-time collusion, defenders can apply various transformations or filters on inputs and outputs to restrict the information that \adv can infer through interactions with $\model$. More broadly, combining complementary defenses that mitigate different risks can further reduce collusion potential~\cite{duddu2024combining}.}

\noindent\change{\textbf{\underline{Limitations and Future Work:}} We identify the following limitations and directions for future work:
\begin{itemize}[leftmargin=*,topsep=0pt,itemsep=0pt]

    \item Our framework focuses on correlations rather than causation, since the limited interpretability of model behavior and the poor understanding of factor--attack dynamics make it difficult to causally identify enabling factors. As a result, our framework is conceptual and the proposed guideline is heuristic in nature, similar to Duddu et al.~\cite{duddu2024sok}. Developing a principled guideline with causal analysis is an important direction for future work.

    \item Our evaluation empirically validates the heuristic guideline without providing theoretical guarantees. We evaluate the framework by (a) demonstrating that it explains collusion in prior work and (b) empirically validating it on five illustrative case studies. However, the evaluation remains limited in establishing the general effectiveness of the guideline. Providing theoretical justification for different collusion settings and their corresponding conditions is an important future research direction.

    \item The scope of attacks explored in this work is limited to well-studied security, privacy, and fairness risks with sufficient prior literature to establish correlations and validate the guideline. Additional attacks are possible, as discussed in \S\ref{sec:discussion} under ``Framework and Requirements: \ref{comprehensive} Comprehensive''. However, further empirical analysis is required to establish the corresponding correlations and apply our guideline.

    \item Although not empirically evaluated, we analyzed prior work spanning different model types (e.g., generative models), settings (e.g., federated learning and retrieval-augmented generation), and modalities (e.g., text, images, and tabular data). This suggests that the identified factors may generalize across diverse models and settings. However, our work does not provide a comprehensive evaluation of such generalization, which we leave for future work. Additionally, other generative-model-specific factors, currently excluded due to unexplored attack correlations, can be naturally incorporated into the framework.
    
\end{itemize}}

\noindent\textbf{\underline{Summary:}}
Existing work lacks a systematic framework for exploring colluding adversaries in ML pipelines. We introduce the \emph{first framework} by identifying factors which enable collusions.
We propose a guideline to identify potential collusion using the factors, explain prior work, conjecture about unexplored ones, and empirically validate five such cases.
We then study how adversaries' characteristics influence the potential for collusion, based on which we update our guideline.

%% file: 00ack.tex
% \section*{Statement of Contributions}

% Asim and Lipeng contributed to Tables \ref{tab:trteeval} and \ref{tab:teeval}, respectively, and assisted with systematizing some papers. They were not involved in writing the paper or designing and validating the framework.
% Vasisht proposed the idea, developed the framework, conducted most of the systematization, and wrote the paper under Asokan’s supervision.

\subsection*{Acknowledgments}

\noindent This work is supported in part by the Wallenberg Visiting Professor Program, the Natural Sciences and Engineering Research Council of Canada (grant number RGPIN-2026- 04826), and the Government of Ontario (RE011-038). Vasisht and Lipeng are supported by David R. Cheriton Scholarship. 
Vasisht is supported by an IBM PhD Fellowship and Mastercard's Cybersecurity and Privacy Excellence Graduate Scholarship. Views expressed in the paper are those of the authors and do not reflect the position of the funding agencies.
The authors thank Anudeep Das, Adam Caulfield, Prach Chantasantitam, and Yihan Wang for their feedback.

%% file: 9appendix.tex
\appendix
\section*{Appendix}

\section{Streamlined Threat Model}\label{sec:streamlined}

We represent an adversary model as \threatModel which is a tuple of sets representing different characteristics of \adv.
To define the threat model, we identify various \adv's characteristics: objectives, capabilities, knowledge, and access to $\model$ and $\dtrain$. Formally, we represent the threat model as a tuple of seven sets (denoted as $\overline{\text{<.>}}$), corresponding to the questions below:
\begin{enumerate*}[leftmargin=*, label={(\roman*)}]
    \item what role(s) does \adv play? (\text{\advSet})
    \item what are \adv's objectives? (\text{\objectiveSet})
    \item what are \adv's capabilities? (\text{\capabilitySet})
    \item what is \adv's interaction with $\model$? (\text{\interactionSet})
    \item what attack optimizations can \adv use? (\text{\optimizationSet})
    \item what is \adv's knowledge about $\model$? (\text{\knowledgeModelSet}); and
    \item what is \adv's knowledge about data? (\text{\knowledgeDataSet})
\end{enumerate*}
Given these characteristics, we denote a threat model \threatModel describing \adv as a tuple of the seven sets:
\[
\text{\threatModel} =
\left(
\begin{array}{l}
    \text{\objectiveSet}, 
    \text{\advSet},
    \text{\capabilitySet},
    \text{\optimizationSet},
    \text{\knowledgeModelSet}, 
    \text{\interactionSet}, 
    \text{\knowledgeDataSet}
\end{array}
\right)
\]
The elements of each of the seven sets constituting \threatModel are drawn from their respective spaces of all possible values. We denote these spaces with the first letter capitalized: 
$\overline{\textbf{\texttt{Obj}}}$ as the space for $\text{\objectiveSet}$, 
$\overline{\textbf{\texttt{Role}}}$ for $\text{\advSet}$, 
$\overline{\textbf{\texttt{Cap}}}$ for $\text{\capabilitySet}$, 
$\overline{\textbf{\texttt{Opt}}}$ for $\text{\optimizationSet}$, 
$\overline{\textbf{\texttt{K}}_{\textbf{\texttt{M}}}}$ for $\text{\knowledgeModelSet}$, 
$\overline{\textbf{\texttt{I}}}$ for $\text{\interactionSet}$, and 
$\overline{\textbf{\texttt{K}}_{\textbf{\texttt{D}}}}$ for $\text{\knowledgeDataSet}$.
We describe the possible values in each space characterizing \adv.
% and summarize them in Table~\ref{tab:summary}.
%
This representation also allows us to define an operation on the tuples to combine adversary models for capturing colluding adversaries.
We outline each set and their possible values, provide an overview of the ML pipeline and involved entities (Figure~\ref{fig:overview}).

\input{figures/fig_overview}
\noindent\textbf{\underline{Role of \adv (\advSet)}} specifies who \adv is. 
This includes data owner (\dataOwner), data provider (\dataProvider), code provider (\codeProvider), model provider (\modelProvider), model owner (\modelOwner), model trainer (\modelTrainer), service provider (\serviceProvider), and client (\client).

\noindent\ding{182}\xspace\textbf{Data Owner} (\dataOwner) can share data taken from several sources (e.g., scraping from the Internet, personal data, software repositories, etc.). This is later used as $\dtrain$ (see \ding{183}) or $\dtest$ (see \ding{188}). 
As \adv, \dataOwner can add poisons/backdoors to $\dtrain$ to degrade utility or give \adv-chosen outputs.

\noindent\ding{183}, \ding{188}\xspace\textbf{Data Provider} (\dataProvider) processes data from data owners to get $\dtrain$ (for training) and $\dtest$ (for evaluation). 
As \adv, \dataProvider can add poisons or backdoors to $\dtrain$ for degrading utility or giving \adv-chosen outputs.

\noindent\ding{184}\xspace\textbf{Code Provider} (\codeProvider) includes third-party libraries (e.g., PyTorch, or code repositories on GitHub) used for training and evaluating $\model$. 
As \adv, \codeProvider can inject malicious code to influence $\model$'s behavior (e.g.,~\cite{blindBackdoors}).

\noindent\ding{185}\xspace\textbf{Model Provider} (\modelProvider) shares the architecture of $\model$ or pre-trained model (e.g., HuggingFace), which can be downloaded for further use (e.g., fine-tuning).
As \adv, \modelProvider can share a backdoored $\model$ (e.g.~\cite{archBackdoor}).

\noindent\ding{186}\xspace\textbf{Model Trainer} (\modelTrainer) uses an architecture or a pre-trained model from \modelProvider, source code from \codeProvider, and $\dtrain$ from \dataProvider, to train or fine-tune $\model$. 
As \adv, \modelTrainer can tamper the training configuration to influence $\model$'s behavior (e.g.,~\cite{blindBackdoors}).

\noindent\ding{187}\xspace\textbf{Model Owner} (\modelOwner) owns the trained $\model$ from \modelTrainer, and gives to \serviceProvider for deployment.
As \adv, \modelOwner can tamper $\model$ to influence its behavior.

\noindent\ding{189}\xspace\textbf{Service Provider} (\serviceProvider) offers the trained $\model$ as a service with an API to send queries and receive outputs. 
As \adv, \serviceProvider mounts various inference attacks (\mia, \aia, \dia, \datarecon) to leak information in $\dtrain$, measure its bias (\disc), or deploy a tampered model.

\noindent\ding{190}\xspace\textbf{Client} (\client) as \adv can mount various privacy attacks (\mia, \aia, \dia, \datarecon), steal $\model$'s functionality (\modelext), and measure its bias (\disc).
Alternatively, \client can download a pre-trained $\model$ (e.g., language models from Hugging Face) and query it locally. In this case, \modelext does not apply, as \client already has full access to $\model$.
\[
\text{\advSet} = \left\{
    \begin{array}{l}\text{\dataOwner}, \text{\dataProvider}, \text{\codeProvider}, \text{\modelProvider}, \\\text{\modelTrainer}, \text{\modelOwner}, \text{\serviceProvider}, \text{\client}
\end{array}
\right\}
\]

\noindent\textbf{\underline{\adv's Objectives (\objectiveSet)}} includes the risks from \Snospace\ref{sec:background}.
\[
\text{\objectiveSet} = \left\{
\begin{array}{l}
    \text{\evasion}, \text{\poison}, \text{\backdoor}, \text{\mia}, \text{\aia},\\
    \text{\dia}, \text{\datarecon}, \text{\modelext}, \text{\disc}
\end{array}
\right\}
\]

\noindent\textbf{\underline{\adv's Capability (\capabilitySet)}} specifies whether or not \adv deviates from the computation prescribed for their role. 
This includes \textbf{honest-but-curious} (\passiveAdv) or \textbf{malicious} (\activeAdv), which have also been called ``passive'' and ``active'' respectively (e.g.,~\cite{nasr2019comprehensive,melis2019exploiting}). 
\passiveAdv cannot deviate from prescribed procedure, but can observe (and use) all visible data. \activeAdv can deviate to do arbitrary computations on top.
\[\text{\capabilitySet} = \{\text{\passiveAdv}, \text{\activeAdv}\}\]

\noindent\textbf{\underline{Optimization (\optimizationSet)}} indicates the type of attack optimization performed by \adv. 
An \textbf{adaptive \adv} (\adaptive) modifies their subsequent attack queries to $\model$, based on the feedback from $\model$'s previous outputs\footnote{\textbf{Terminology:} ML literature misuses ``adaptive'' to describe \adv aware of defenses~\cite{tramer2020adaptive,feng2023stateful,lukas2024leveraging,FenauxSoK}, whereas in security literature, such knowledge is assumed by default. 
We call adversaries without this knowledge ``na\"{i}ve''. 
Following standard security terminology~\cite{adaptivesecurity,adaptiveCrypto1}, we define ``adaptive'' adversaries as those who update their attacks based on previous responses, and assume \adv knows the defenses by default.}.
Alternatively, \textbf{non-adaptive} \adv  (\nonAdaptive) use a fixed set of attack queries or optimize them locally without feedback from $\model$.
\[\text{\optimizationSet} =\{\text{\adaptive}, \text{\nonAdaptive}\}\]

\noindent\textbf{\underline{Knowledge of Model (\knowledgeModelSet)}} includes \textbf{whitebox} (\whitebox), \textbf{blackbox} (\blackbox), \textbf{graybox} (\graybox), and \textbf{no access} (\noAccess). 
For both \blackbox and \graybox, we have \textbf{full predictions} (\fullPrediction), \textbf{top-k predictions} (\topK), and \textbf{hard labels} (\hardLabel).
\begin{itemize}[leftmargin=*]
    \item \textbf{Whitebox} (\whitebox): \adv has full knowledge of $\model$'s architecture, parameters, training process and training configuration (e.g., downloaded publicly available pre-trained models). Since \adv has access to $\model$'s parameters, they can compute gradients and intermediate activations.
    \item \textbf{Graybox} (\graybox): \adv knows the architecture and training configuration but not the parameters.
    \item \textbf{Blackbox} (\blackbox): \adv does not know $\model$'s architecture, parameters, training process and configuration.
    For both \blackbox and \graybox, we have the following cases:
    \begin{itemize}[leftmargin=*]
        \item \textbf{Hard-Label} (\hardLabel): \adv only receives the final classification for a given input, such as malware classifiers producing labels (e.g., malware or not).
        \item \textbf{Top-K Predictions} (\topK): \adv obtains top-k confidence scores from the prediction vector ($1 \leq k \leq N$) with $N$ classes (e.g., ImageNet models output scores for a subset of classes).
        \item \textbf{Full Predictions} (\fullPrediction): \adv receives confidence scores for all classes (e.g., in high-stakes ML models like credit approval).
    \end{itemize}
    \item \textbf{No access} (\noAccess): \adv cannot query $\model$ and can access the outputs for some inputs.
\end{itemize}
\[
\text{\knowledgeModelSet} = \left\{
\begin{array}{l}
    \text{\noAccess},\ (\text{\blackbox},\ \text{\graybox})\\
    \times (\text{\hardLabel},\ \text{\topK},\ \text{\fullPrediction}), \text{\whitebox},
\end{array}
\right\}
\]
We use (\blackbox, \text{\graybox}) $\times$ (\hardLabel, \topK, \text{\fullPrediction}) to denote their six possible combinations.
\whitebox corresponds to \fullPrediction (\whitebox.\fullPrediction), which we omit for brevity.

\noindent\textbf{\underline{Interaction (\interactionSet)}} indicates how \adv interacts with $\model$:
\begin{enumerate*}[leftmargin=*,label={(\roman*)},itemjoin={;\xspace}]
    \item \textbf{no access (\noAccess)}
    \item \textbf{one-shot (\oneShot)} where \adv can only query once (e.g., evading a malware classifier on a target machine)
    \item \textbf{k-Shot (\kShot)} where \adv can send $k$ queries to $\model$ (e.g., in case of budget constraints)
    \item \textbf{unlimited (\unlimited)} where \adv has no limit on the number of queries to $\model$ (e.g., with local access to $\model$).
\end{enumerate*}
\[\text{\interactionSet} =\{\text{\noAccess}, \text{\oneShot}, \text{\kShot}, \text{\unlimited}\}\]

\noindent\textbf{\underline{Knowledge of Data (\knowledgeDataSet)}} indicates the assumptions about \adv's auxiliary data $\daux$ for designing ($\dauxtrain$) and evaluating ($\dauxtest$) the attack. 
For $\dauxtrain$, we indicate whether \adv has \textbf{no access} (\noAccess), \textbf{no overlap} (\noOverlap), \textbf{partial overlap} (\partialOverlap) or \textbf{full overlap} (\fullOverlap) with $\dtrain$. For $\dauxtest$, we indicate if \adv has ground truth: \textbf{non-blind} (\noBlind) when available, and \textbf{blind} (\blind) otherwise (e.g., unavailable as labeling is expensive).
\begin{itemize}[leftmargin=*]
    \item \textbf{No Access} (\noAccess): Some \adv in ML pipeline will not have access to $\dtrain$ (e.g., \codeProvider or \modelProvider).
    \item \textbf{No Overlap} (\noOverlap): \adv knows the underlying distribution of $\dtrain$ from which \adv samples $\dauxtrain$ but has no overlap with $\dtrain$.
    Also, $\dauxtrain$ can include natural data or synthetically generated data.
    \item \textbf{Partial Overlap} (\partialOverlap):
    $\dauxtrain$ has partial overlap with $\dtrain$ (e.g., part of $\dtrain$ is public).
    \item \textbf{Complete Overlap} (\fullOverlap): $\dauxtrain$ is the same as $\dtrain$ including features and labels (e.g., $\dtrain$ is public for GPT2). 
\end{itemize}
We denote the possible combinations as \{\noOverlap, \partialOverlap, \text{\fullOverlap}\} $\times$ \{\noBlind, \text{\blind}\}.
\[
\text{\knowledgeDataSet} = \left\{
\begin{array}{l}
    (\text{\noOverlap},\ \text{\partialOverlap},\ \text{\fullOverlap}) \\
    \times\ (\text{\noBlind},\ \text{\blind})
\end{array}
\right\}
\]
% Given the different values for each set, there are a large number of possible adversary models. However, not all adversary models are valid since some of the values cannot be applied together. 
% We describe the constraints on \adv's characteristics to identify valid adversary models (in Table~\ref{tab:trends}).

% \input{tables/tab_adv_desc}

\section{Evaluating Streamlined Threat Model}
% We now discuss how our proposed streamlined adversary model is comprehensive and extensible.

\noindent\textbf{Comprehensive:} We show that our threat model can represent prior work covering different risks, model types, settings. 
We show this in our open-sourced artifact where we specify \adv's characteristics covering individual risks and collusions from the literature survey\footnote{Artifact: \url{https://github.com/ssg-research/sok-collusion}\label{artifact}}. 
% We clarify our methodology for collecting papers below. 
% 
% \noindent\emph{Methodology for Collecting Papers on Individual Risks:}
We survey some papers to cover a variety of model types and settings, to \emph{illustrate} that our streamlined threat model is comprehensive.
\begin{itemize}[leftmargin=*]
    \item To cover different model types, we searched for the following keywords: \emph{language models, audio models (automated speech recognition), graph models, image classifiers, vision language models, reinforcement learning, diffusion models, text-to-image models, generative adversarial networks, and variational autoencoders}.
    \item To cover different settings, we searched for the following keywords: \emph{MLaaS, transfer learning, online learning, federated learning, and vertical federated learning}.
     % \footnote{For federated learning, we mark the participants who are also model trainers, as data owners, while the aggregating server as the model owner.}
    \item We selected papers from top tier security and privacy (e.g., USENIX, NDSS, S\&P, and CCS), machine learning (e.g., ICML, ICLR, NeurIPS, AAAI), and computer vision (e.g., CVPR, ICCV, ECCV) venues.
    \item Overall, we collected 95 papers for \evasion, 53 for \poison/\backdoor, 29 for \modelext, 70 for \mia, 32 for \aia, 13 for \dia, and 74 for \datarecon. 
    % As \disc does not have specific attacks and corresponding threat models, we did not cover its papers.
\end{itemize}

\noindent\textbf{Extensible:} To add a new role for \adv, we update \advSet and specify \adv's characteristics (\capabilitySet, \optimizationSet, \interactionSet, \knowledgeModelSet and \knowledgeDataSet). For new risks, we update objective set (\objectiveSet) to define who can perform the attack, describe the optimizations (\optimizationSet), type of interaction (\interactionSet), and the required knowledge (\knowledgeModelSet, \knowledgeDataSet).

%% file: figures/fig_overview.tex
\pgfdeclarelayer{background}
\pgfdeclarelayer{foreground}
\pgfsetlayers{background,main,foreground}
\tikzstyle{defense} = [rectangle,  minimum width=1.4cm, minimum height=0.5cm, text centered, draw=black]

\begin{figure}[!htb]
\begin{center}
\resizebox{\columnwidth}{!}{ 
\begin{tikzpicture}[line width=1pt]

\node (dataProvider) [align=center] {\footnotesize \textbf{\ding{183} Data Provider}\\\footnotesize \textbf{(\dataProvider)}};
\node (trdata) [defense, below of=dataProvider, align=center, minimum width=1.5cm] {{\small \faDatabase}\\\footnotesize (Training Data)};

\begin{scope}[on background layer]
    \node (dataProvider2) [fit=(trdata) (dataProvider), fill= green!8, rounded corners, draw=black, dashed, inner sep=.08cm] {};
\end{scope}

\node (data1) [database, above of=dataProvider2, xshift=-1cm,yshift=0.75cm, fill= cyan!8, align=center, database radius=0.2cm,database segment height=0.1cm] {};
\node (data2) [database, right of=data1, xshift=-0.3cm, fill= cyan!8, align=center, database radius=0.2cm,database segment height=0.1cm] {};
\node (data3) [database, right of=data2, xshift=-0.3cm, fill= cyan!8, align=center, database radius=0.2cm,database segment height=0.1cm] {};
\node (data4) [right of=data1, xshift=1.5cm, align=center] {\footnotesize \textbf{$\cdots \cdots \cdots \cdots$}};
\node (data5) [database, right of=data4, xshift=0.2cm, fill= cyan!8, align=center, database radius=0.2cm,database segment height=0.1cm] {};
\node (data6) [database, right of=data5, xshift=-0.3cm, fill= cyan!8, align=center, database radius=0.2cm,database segment height=0.1cm] {};
\node (data7) [database, right of=data6, xshift=-0.3cm, fill= cyan!8, align=center, database radius=0.2cm,database segment height=0.1cm] {};

\node (dataOwners) [right of=data7, xshift=1.3cm, align=center] {\footnotesize \textbf{\ding{182} Data Owners (\dataOwner)}};

\begin{scope}[on background layer]
    \node (dataOwners2) [fit=(data1) (data2) (data3) (dataOwners), fill= green!8, rounded corners, draw=black, dashed, inner sep=.08cm] {};
\end{scope}

\draw[->, ultra thick] ([xshift=-3.47cm]dataOwners2.south) -- (dataProvider.north);

% \draw[->, ultra thick] (data2.south) -- (dataProvider.north);
% \draw[->, ultra thick] (data3.south) -- ([xshift=0.2cm]dataProvider.north);

\node (codeProvider) [right of=dataProvider2, yshift=0.5cm, xshift=2cm,align=center] {\footnotesize \textbf{\ding{184}\xspace Code Provider}\\\footnotesize \textbf{(\codeProvider)}};

\node (code) [defense, below of=codeProvider, align=center, minimum width=1.5cm] {{\small \faCodeFork}\\\footnotesize (Code Library)};

\begin{scope}[on background layer]
    \node (codeProvider2) [fit=(codeProvider) (code), fill= black!8, rounded corners, draw=black, dashed, inner sep=.08cm] {};
\end{scope}

\node (modelProvider) [right of=codeProvider,xshift=2.4cm, align=center] {\footnotesize \textbf{\ding{185}\xspace Model Provider}\\\footnotesize \textbf{(\modelProvider)}};

\node (pretrained) [defense, below of=modelProvider, align=center, yshift=0.05cm, xshift=1cm, minimum width=1.5cm] {\footnotesize Pre-trained\\\footnotesize Model};

\node (modConfig) [draw, below of=modelProvider, align=center, xshift=-0.8cm] {{\small \faFileText\xspace} {\footnotesize(Model} \\ {\footnotesize Description)}};

\begin{scope}[on background layer]
    \node (modelProvider2) [fit=(pretrained) (modelProvider) (modConfig), fill= blue!8, rounded corners, draw=black, dashed, inner sep=.08cm] {};
\end{scope}

\node (modelTrainer) [below of=dataProvider, yshift=-1.5cm, xshift=0.2cm, align=center] {\footnotesize \textbf{\ding{186} Model Trainer}\\\footnotesize \textbf{(\modelTrainer)}};

\node (trConfig) [draw, below of=modelTrainer, align=center, yshift=0.25cm] {{\small \faFileText}\hspace{0.05cm} {\footnotesize (Train Config.)}};

\node (training) [rounded rectangle, draw, right of=trConfig, xshift=4.5cm, align=center, minimum width=5cm, minimum height=0.5cm] {{\footnotesize \textbf{Training}} {\small \faGears} };

\begin{scope}[on background layer]
    \node (training2) [fit=(modelTrainer) (trConfig) (training), fill= blue!8, rounded corners, draw=black, dashed, inner sep=0.2cm] {};
\end{scope}

\draw[->, ultra thick] (dataProvider2.south) -- ([xshift=-0.75cm]training.north);
\draw[->, ultra thick] (codeProvider2.south) -- ([xshift=-0.2cm]training.north);
\draw[->, ultra thick] (modelProvider2.south) -- ([xshift=0.75cm]training.north);
% \draw[->, ultra thick] (modConfig.south) -- ([xshift=0.25cm]training.north);
\draw[->, ultra thick] (trConfig.east) -- (training.west);

\node (model) [defense, below of=training, align=center, yshift=-0.65cm, xshift=-3cm] {\footnotesize Trained\\\footnotesize Model};

\node (modelOwner) [right of=model, minimum width=2cm,align=center, yshift=-0.25cm, xshift=1.25cm] {\footnotesize \textbf{\ding{187} Model Owner}\\ \footnotesize \textbf{(\modelOwner)}};

\node (metrics) [right of=model, align=center,xshift=3.85cm] {\footnotesize \textbf{Evaluation}\\\footnotesize \textbf{Metrics}};

\begin{scope}[on background layer]
    \node (modelOwner2) [fit=(model) (modelOwner), fill= blue!8, rounded corners, draw=black, dashed, inner sep=.15cm] {};
\end{scope}

\draw[->, ultra thick] ([xshift=-2cm]training.south) -- (model.north);

% % \node (tedata) [defense, above of=modelOwner, align=center, yshift=0.85cm] {\footnotesize Evaluation\\\footnotesize Dataset};

\node (dataProvider3) [left of=model, xshift=-1.7cm, yshift=0.2cm, align=center] {\footnotesize \textbf{\ding{188} Data Provider}\\\footnotesize \textbf{(\dataProvider)}};

\node (tedata) [defense, below of=dataProvider3, yshift=0.1cm, align=center, minimum width=1.5cm] {{\small \faDatabase}\\\footnotesize (Evaluation Data)};

\begin{scope}[on background layer]
    \node (dataProvider4) [fit=(tedata) (dataProvider3), fill= green!8, rounded corners, draw=black, dashed, inner sep=.08cm] {};
\end{scope}

\draw[->, ultra thick, blue] ([yshift=-0.3cm]dataProvider3.east) -- ([yshift=-0.1cm]model.west);
\draw[->, ultra thick, blue] ([yshift=0.2cm]model.east) -- ([yshift=0.2cm]metrics.west);
\draw[->, ultra thick]
  (dataOwners2.west) -- ++(-0.25,0) |- (dataProvider3.west);

% \draw[->, ultra thick] ([xshift=-3.5cm]dataOwners2.west) -| (dataProvider4.west);

% \draw[->, ultra thick, blue] (model.south) -- (metrics.north);

\node (depmodel) [defense, below of=model, yshift=-1cm, align=center] {\footnotesize Deployed\\\footnotesize Model};

\node (serviceProvider) [left of=depmodel, xshift=-1.5cm, align=center] 
{\footnotesize\shortstack{\textbf{\ding{189} Service Provider}\\(\textbf{\serviceProvider)}}};

\begin{scope}[on background layer]
    \node (serviceProvider2) [fit=(serviceProvider) (depmodel), fill= orange!8, rounded corners, draw=black, dashed, inner sep=.1cm] {};
\end{scope}

\draw[->, ultra thick] (model.south) -- (depmodel.north);

\node (inp) [right of=depmodel, yshift=0.15cm, xshift=1.5cm,minimum width=0.1cm] {\footnotesize \textbf{Input}};
\node (out) [right of=depmodel, yshift=-0.15cm, xshift=1.5cm, minimum width=0.1cm] {\footnotesize \textbf{Output}};
\node (client) [right of=depmodel, align=center, xshift=3.25cm] 
{\footnotesize\textbf{\ding{190} Client} (\client)};

\draw[->, ultra thick] (inp.west) -- ([yshift=0.15cm]depmodel.east);
\draw[->, ultra thick] ([yshift=-0.15cm]depmodel.east) -- (out.west);

\begin{scope}[on background layer]
    \node (client) [fit=(client) (inp) (out), fill= yellow!8, rounded corners, draw=black, dashed, inner sep=.1cm] {};
\end{scope}

\end{tikzpicture}
}
\end{center}
\caption{\textbf{ML Pipeline.} Raw data from data owners (\dataOwner) is aggregated by a data provider (\dataProvider) and processed into $\dtrain$ and $\dtest$. A model trainer (\modelTrainer) uses architecture from a model provider (\modelProvider), code from a code provider (\codeProvider), and a training configuration to train or fine-tune $\model$. The resulting $\model$ is owned by a model owner (\modelOwner), who evaluates it using $\dtest$. Finally, a service provider (\serviceProvider) deploys the model for clients (\client).}
\label{fig:overview}
\end{figure}